%% file: ms.tex
\documentclass{article}
\usepackage[utf8]{inputenc}

\usepackage{times}
\usepackage{geometry}
\usepackage{amsmath}
\usepackage{amssymb}
\usepackage{amsthm}
\usepackage{bbm}
\usepackage{array, multirow, graphicx}
\usepackage{booktabs}
\usepackage{longtable, wrapfig, colortbl, pdflscape, tabu, threeparttable}
\usepackage{hyperref}
\usepackage{color}
\usepackage[table,x11names]{xcolor}
\hypersetup{colorlinks=true, urlcolor = DodgerBlue4, citecolor = black}
\usepackage{setspace}
\usepackage{subfig}
\usepackage[english]{babel}
\usepackage{natbib}
\usepackage{float}
\usepackage{wasysym}   
\usepackage{comment}
\usepackage{enumitem}
\usepackage{import}
\usepackage{arydshln}
\bibpunct{(}{)}{;}{a}{,}{,}

\newtheorem{assumption}{Assumption}

\newtheorem{hypothesis}{Hypothesis}
\newtheorem*{hypothesis*}{Hypothesis}

\newtheorem{proposition}{Proposition}
\newtheorem{result}{Result}

\newcommand\todonote[1]{}

\usepackage{amsmath}
\DeclareMathOperator*{\argmax}{arg\,max}

\usepackage{tcolorbox}

\newcommand{\vect}[1]{\boldsymbol{#1}}

\title{Arbitraging Narrow Bracketers \thanks{A previous version of this paper was circulated under the title "Narrow Bracketing in Work Choices". The studies were pre-registered in the AEA RCT Registry with ID 0003412 (see \url{https://doi.org/10.1257/rct.3412}) and OSF (see \href{https://osf.io/5q7wn/}{https://osf.io/5q7wn/}). The data and analysis are hosted on GitHub at \url{https://github.com/MarcKaufmann/narrow-bracketing-in-work-choices}.}}

\author{
  Francesco Fallucchi\\
  University of Bergamo\thanks{\texttt{ francesco.fallucchi@unibg.it},  Senior Assistant Professor, Department of Economics, Via dei Caniana 2, 24127 Bergamo, Italy.}
  \and
  Marc Kaufmann\\
  Central European University (CEU)\thanks{Corresponding author: \texttt{kaufmannm@ceu.edu}, Associate Professor, Department of Economics and Business, Central European University, Quellenstrasse 51-55, 1100 Vienna, Austria.}
}

\date{\today}

\usepackage{natbib}
\usepackage{graphicx}

\begin{document}

\maketitle

\begin{abstract}
  Many important economic outcomes result from the combined effects of several choices, so the best option is not determined from each choice in isolation, but depends on how each choice alters total outcomes. We formally show that narrow bracketing --- treating choices in isolation --- can be distinguished from broad bracketing --- combining the choices --- if and only if there exist price variation across context: there is some bundle for which a person is willing to pay more in one choice than in another. In this case, a narrow bracketer can be arbitraged, buying the bundle when it is expensive and selling when it is cheap in simultaneous choices. We design and run two experiments to identify bracketing from price variation. In a between-subjects design where we vary the amount of work to generate price variation, we reject broad bracketing and fail to reject narrow bracketing. In a within-subject design we directly test bracketing by attempting to arbitrage our participants. For price variation coming from varying amounts of money, 50.3\% of subjects are classified as narrow bracketers, and only 14.6\% as broad bracketers, the remainder being inconsistent with both. This changes for price variation coming from violations of expected utility --- 13.6\% narrow vs 46.3\% broad --- and of the weak axiom of revealed preference --- 26.3\% narrow vs 38.1\% broad. 

\end{abstract}

\newpage

\section{Introduction}\label{sec:intro}

\import{./}{new-introduction.tex}

\section{Conceptual Framework of Narrow Bracketing}\label{sec:framework}

\import{./}{new-framework.tex}

\section{Study 1: Testing for bracketing of work choices}\label{sec:design1}

\import{./}{experiment-1-design-resubmission.tex}

\section{Study 1: Results}\label{sec:results1}

\import{./}{data/experiment-results.tex}

\section{Study 2: Testing for choice bracketing across domains}\label{sec:design2}

\import{./}{experiment-2-design-resubmission.tex}

\section{Study 2: Results}\label{sec:results2}

\import{./}{experiment-2-results.tex}

\section{Conclusion}\label{sec:conclusion}

\import{./}{conclusion.tex}

\section{Acknowledgements}

We thank David Freeman, Gerg\H{o} Hajdu, Holger Herz, Mats Köster, Botond K\H{o}szegi, Tomy Lee, Robert Lieli, Sergey Lychagin, Arieda Muço, Daniele Nosenzo, Ádám Szeidl, Georg Weizsäcker, seminar participants at CEU, Unversity of Portsmouth, University of Trento, BEEN 2020 (University of Verona), ESA 2020, ASFEE 2021 (Dijon), YEM 2022 (Brno), ASFEE 2022 (Lyon), and FUR 2022 (Ghent) conferences for helpful comments.

\newpage

\bibliographystyle{plainnat}
\bibliography{references}

\newpage

\appendix

\import{./}{appendix.tex}

\end{document}

%% file: new-introduction.tex
Many of the most important economic outcomes --- such as savings, health, careers, and relationships --- result from countless interdependent decisions. In the work domain, such cumulative decision making is bound to become even more important as remote, flexible, and online work becomes more popular \citep{farrell2018online,farrell2019evolution}. In order to obtain the best outcomes in these circumstances, a person must, in every decision, at least take into account cumulative outcomes so far. Whenever a person fails to do so, she is engaging in narrow bracketing (\cite{read1999choice}). The prior literature focuses on nonwork choices, such as monetary and lottery choices (\cite{tverskyKahneman1981framing}, \cite{redelmeier1992framing}, \cite{rabin2009narrow}, \cite{koch2019correlates}, \cite{vorjohann2020referenceDependentChoiceBracketing}, and \cite{dekelGoldwaterLovalloBurns2021choiceBracketingRiskAggregation}); social choices \cite{exley2018equity}; or both of these and induced preferences (\cite{ellisFreeman2020revealingBracketing}). Broad bracketing in these situations is either complicated (when combining lotteries), or not necessarily the relevant taget (when making social choices). In contrast, we investigate narrow bracketing in economically essential work choices, where accounting for these outcomes is simple and relevant.

We contribute to the literature on bracketing by formalizing a measurable notion of price variation across choices that is key to identifying narrow bracketing and apply it to identify narrow bracketing across two experiments. First, we show that every narrow bracketer can be arbitraged across choices with different buying and selling prices: they purchase a good when it is expensive in one choice only to sell it when it is cheap in another. We characterize the set of preferences under which such arbitrage opportunities do not arise due to equal buying and selling prices, in which case narrow bracketers behave optimally and are observationally equivalent to broad bracketers. Second, building on this theoretical foundation, we design and conduct two experiments to identify price variations arising from different sources --- from nonlinear preferences, violations of expected utility, or inconsistencies with the weak axiom of revealed preference (WARP) --- and use these variations to classify individuals as narrow bracketers, broad bracketers, or inconsistent with both.

To illustrate buying and selling prices, consider a person making choices over money and hours of work. Suppose that this person has a marginal disutility of \$$5$ per hour of work for the first hour of work, and a marginal disutility of \$$10$ per hour for every additional hour. Then this person chooses 1 hour of work for \$$6$ over no work for no pay --- or in other words, they sell 1 hour of work for \$$6$. This person also chooses 1 hour of work for no pay over 2 hours of work for \$$9$ --- i.e., they refuse to sell 1 \emph{extra} hour of work (the difference between the two options) for \$$9$, which we refer to as \emph{buying} one hour of work for \$$9$. Defining buying and selling of 1 extra hour of work in this way, we find that the highest buying price for 1 extra hour of work is \$$10$, and the lowest selling price is \$$5$
Now consider a person who is given both of these choices simultaneously and who will receive their combined outcome. Narrow bracketing requires that the person optimizes each choice separately, so a narrow bracketer would choose the same way as if each choice was the only choice. Therefore, they would sell 1 hour of work for \$$6$ in the first choice, and buy 1 hour of work for \$$9$ in the second choice, even though switching both choices by buying 1 hour for \$$6$ in selling 1 hour of work for \$$9$, would have earned them \$$3$ more for the same total amount of 2 hours of work. A broad bracketer would optimize the sum of the choices instead, and thus choose between the following total outcomes: 3 hours of work for \$$15$, 2 hours for \$$9$, 2 hours for \$$6$, or 1 hour for no money. Therefore, they would never choose 2 hours for \$$6$ over 2 hours for \$$9$ and so cannot be arbitraged by offering simultaneous choices. 

This example highlights our first result in Section \ref{sec:framework}, which shows that whenever there exists some bundle for which the highest buying price exceeds the lowest selling price, then a narrow bracketer can be arbitraged by buying high in one choice and selling low in another. More surprisingly, if for every bundle $X$ the highest buying price equals the lowest selling price, denoted $P(X)$, then a narrow bracketer behaves identically to a broad bracketer, and thus we cannot identify bracketing. We show that this occurs if and only if the choices can be rationalized by preferences --- so that they satisfy the weak axiom of revealed preference (WARP) --- and that $P(\cdot)$ provides a utility representation for these preferences, satisfying additivity: $P(X + Y) = P(X) + P(Y)$. Additivity of $P(\cdot)$ implies that from each choice set, the person chooses the bundle with the highest price, which implies in particular that the utility over amounts is linear in quantities and utility over lotteries is linear in probabilities, and so satisfies expected utility. For people with such preferences, narrow bracketing is fundamentally unidentified, and so in order to identify bracketing, we must find some bundle for which the highest buying price exceeds the lowest selling price. This is guaranteed to be the case if we observe a violation of WARP, of expected utility, or of linear utility.

In our first main experiment described in Section \ref{sec:results1}, a between-subjects design conducted on Amazon Mechanical Turk recruiting 716 subjects, we generate price variation by varying the amount of tasks participants have to do by default, in order to test a particular form of narrow bracketing often called endowment bracketing. 
Concretely, in our first scenario, some of the participants are asked for the smallest price --- their reservation wage --- for which they are willing to do 15 tasks rather than 0 tasks, while others are asked for their reservation wage for doing 30 rather than 15 tasks. We find that these reservation wages are significantly different, \$2.88 vs. \$2.31 (p-value$<0.001$) respectively, which provides us with the difference in prices needed to identify population-level bracketing. In a third treatment, participants are told on the same page that they have to do a baseline of 15 tasks and then asked for their reservation wage for doing 15 additional tasks. Under this treatment, a broad bracketer would recognize that the decision is equivalent to choosing between 30 tasks instead of 15, whereas a narrow bracketer would perceive it as a choice between 15 tasks and none. We find that the reservation wage elicited is \$2.07 --- 28\% significantly lower than the reservation wage of \$2.88 expected for broad bracketers, but not significantly different from the reservation wage of \$2.31 expected for narrow bracketers, suggesting that work decisions are consistent with narrow bracketing. In a second scenario, where we increase the number of baseline tasks in each treatment by 15, we do not generate enough price variation and, therefore, fail to reject both narrow and broad bracketing.
In Section \ref{sec:design2}, we describe the design of our second study, where we use arbitrage in a within-study design to test for narrow bracketing beyond endowment bracketing. Following closely the structure of our example described above, we first search for pairs of choices that reveal potential for arbitrage: participants are willing to buy work for a higher price in one choice than they are willing to sell in the other. We do so by offering participants choices from three domains, each varying a different feature to identify a potential source that would generate a price difference. First, we look for non-linear utility by varying the amount of tasks; second, we look for violations of WARP by adding extra (unchosen) options to choice sets; and finally, we look for violations of expected utility by varying the probability with which a fixed amount of tasks has to be done. If we identify such pairs of choices, we let participants make this pair of choices simultaneously, as well as make a single choice equal to the sum of both choices, so that a narrow bracketer will throw away money when making the simultaneous choices, but not when making the combined choice.\footnote{Assuming they prefer more money to less, which might not hold in actual data for instance if participants are inattentive.} A broad bracketer, in contrast, will exhibit the same behavior across both scenarios, allowing us to classify an individual as aligning with narrow but not broad bracketing if they discard money in the simultaneous choice; as aligning with broad but not narrow bracketing if their choices match in both the simultaneous and the combined scenarios; or as neither in other cases. 

Our results from this experiment are described in Section \ref{sec:results2}. We conducted it on Prolific, recruiting 572 subjects to make 35 decisions of work linked to the three domains described above. We find that around 60\% of the participants who complete the experiment are identified under our arbitrage condition in at least one of the three settings, and on average we have an identification of 30\% in each setting. Among them, we find that roughly 50\% are identified as narrow bracketers in the linear setting, while less than 15\% are broad bracketers; in the other two settings, the share of broad bracketers is higher than the share of narrow bracketers: 38\% vs. 26\%  for WARP violations and 46\% vs. 14\% for violations of expected utility. The remaining subjects cannot be classified as inconsistent with both narrow and broad bracketing. While our focus is on individuals for whom we can find pairs of choices that ensure price variation, we also offer simultaneous and combined choices to the remaining participants and categorize them, since these extra choices might themselves provide identifying variation. As is to be expected, the majority of these individuals remain unidentified (on average 70\% in each domain, with 39\% of participants unidentified in all domains), as their choices are consistent with WARP, expected utility, and linear utility, and thus with both narrow and broad bracketing. For the remaining subjects, we have findings similar to those among the identified ones.

Our contribution to the literature on narrow bracketing is threefold. First, we provide a characterization for when bracketing is identifiable --- that narrow bracketing can be distinguished from broad bracketing --- and show that when it is the case, we can identify it via an arbitrage test. Such arbitrage was already present in \cite{tverskyKahneman1981framing} and explicitly highlighted in \cite{rabin2009narrow} for the specific case of lotteries over money when absolute risk aversion is varying and in \cite{andreoni_arbitrage_2018} for the case of arbitraging payments for work over time. Our result establishes that such arbitrage arguments can be applied much more broadly and are an identifying feature of narrow bracketing. An advantage of arbitrage tests for narrow bracketing is that they make clear that the behavior is a mistake, since the person throws away money, whereas more indirect tests --- such as in our first experiment --- cannot rule out that the person made a mistake in the combined choice. We relate our characterization result to the literature at the end of Section \ref{sec:framework}, where we briefly discuss where the identifying variation in prices comes from in earlier tests of bracketing. 

Second, our framework provides direct tests for both narrow and broad bracketing, as well as a test for whether bracketing is identified in a given dataset. Some papers only test (and usually reject) broad bracketing (\cite{rabin2009narrow}, \cite{abeler2017fungibility}, \cite{koch2019correlates}) and thus cannot establish that what people do when they are not bracketing broadly is actually narrow bracketing, other papers do effectively test for both (\cite{kahneman_prospect_1979}, problem 10), but none provide a general framework for testing both. One notable exception is \cite{ellisFreeman2020revealingBracketing} who apply a revealed preference approach to test narrow and broad bracketing. In their framework, conceptually narrow and broad bracketing are only defined for choices consistent with transitive preferences, and so practically one cannot talk about or test bracketing in any choice dataset violationg WARP. Our framework defines bracketing rather as a property of choices and thus extends to situations where people are subject to decoy, range, or reference effects. Moreover, our finding that price variation is essential provides practical guidance to researchers seeking identification.

Finally, our experimental results using the arbitrage test show the value of testing for narrow bracketing across different variations of the choices. Our results for price variation driven by non-linear utility --- most similar to previous studies --- strongly support narrow bracketing. The evidence in favor of narrow bracketing is weaker for price variation driven by violations of expected utility or of WARP. This attenuation may result from our within-subjects design, which could make direct arbitrage between choices more salient, or it may indicate that different variations encourage more deliberate integration of choices, shedding light on the mechanisms underlying narrow bracketing.

We conclude in Section \ref{sec:conclusion}, discussing identification strategies that don't rely on direct (and thus overly salient) arbitrage, and how the identification condition will change when agents make choices from choices with additional structure --- such as choosing how much to buy of one good at once, but making separate choices for each good.

%% file: new-framework.tex
In this section, we consider an agent with a given value function $v$ that is monotonic in money and ask when a person can incur a cost from narrow bracketing when facing two simultaneous choices. We show that narrow bracketing leads to different choices and thus to a cost if and only if there is variation across choice contexts in what we define the buying and selling prices for a bundle of goods. Under such variation, a narrow bracketer can be arbitraged by providing them with two choices, one in which they choose the bundle despite giving up a large amount of money (the price is high), one in which they choose a small amount of money over the bundle (the price is low) --- in other words, they buy high and sell low. 
When these prices are constant across all choice contexts, then a narrow bracketer behaves like a broad bracketer, thus incurs no costs, and it is impossible to identify broad from narrow bracketing. Such constant prices imply that the value function $v$ is consistent with transitive preferences that are additive: they satisfy expected utility, have constant absolute risk aversion, and are linear in each of the goods for two or more goods. 

\subsection{Setup and Definitions}\label{subsec:framework-setup-and-definitions}

\paragraph{Domain of Choice.} An agent faces $i \geq 1$ simultaneous choices from choice sets $S_i$. We will assume that each $S_i$ has at most $N$ options for some fixed but finite $N$. Each option $x \in S_i$ is a bounded, random amount of $n + 1$ goods, the first $n \geq 0$ goods are consumption goods and the $n + 1$st good is money.\footnote{Any good that satisfies the conditions we state in the next paragraph can serve as money.} So each $x$ is a bounded random variable over $\mathbbm{R}^{n + 1}$, which we denote by $\mathbbm{X}$.

\paragraph{Value function.} Let $S$ be the set of possible combined outcomes, i.e., $S = \sum_i S_i$. When choosing directly over final outcomes, the agent's choices are derived by maximizing the value function --- or decision utility --- $v(x|S)$ over final outcomes $x \in S$. Thus under the assumption that the agent makes no mistakes in single choices, this assumes that an agent's utility depends only on the final outcome $x$ and on the set of possible alternatives $S$, but not on whether this was achieved in a single or multiple choices. We assume that $v(\cdot | S)$ is continuous in $x$ and $S$.\footnote{Formally, let $S_0 = S \setminus \{x\}$, then $v(x|S_0 \cup \{x\})$ is continuous in $x$ and $S_0$. } With some abuse of notation, we write $m$ (for \emph{money}) to denote bundles consisting only of sure amounts of money with $m \in \mathbb{R}$.
We further assume monotonicity in money-only choices and non-satiation in money. Monotonicity in money-only choices implies that when $S = \{m_1, \dots, m_n \}$ is a choice between $n$ monetary amounts, then $m_i > m_j \iff v(m_i|S) > v(m_j|S)$. Non-satiation in money holds if adding or subtracting sufficiently large sums of money from an option makes it eventually the only chosen option, or not chosen. Formally, for any $S$ and any $X \in S$, let $S_m \equiv S \setminus \{X\} \cup \{X - m\}$ be the set where $X$ is replaced by $X - m$. Then there is some $m_B$ such that only $X - m$ is chosen from $S_m$ for all $m < m_B$ and some $m_S$ such that $X - m$ is never chosen for any $m > m_S$.\footnote{Note that this allows for local non-monotonicity in money, say, due to strong decoy effects and thus allows for a wide range of effects driven by context or comparison set effects.} 

\paragraph{Narrow and Broad Bracketing.} Consider a person who makes two simultaneous choices from $S_1$ and $S_2$.\footnote{Our definition straightforwardly generalizes to any $i > 1$.} We say that a person \emph{brackets narrowly} if they maximize $v$ in $S_1$ and $S_2$ separately. So they choose $x_i^N$ from $S_i$ satisfying $x_i^N \in \argmax_{x \in S_i} v(x|S_i)$, and thus end up with a combined bundle $z^N = x_1^N + x_2^N$. Similarly, we say that a person \emph{brackets broadly} if they maximize $v$ over the combined choice set $S = S_1 + S_2$: they pick $z^B$ satisfying $z^B \in \argmax_{z \in S} v(z|S)$.

\subsection{Arbitraging Narrow Bracketers}

\paragraph{Buying and Selling Prices} Consider some bundle $X$. We will now define buying and selling prices for $X$ in different choice contexts. Take some $Y$ and $X$ and some $S$ not containing $Y$. Consider a person who chooses from the choice set $S_m = S \cup \{Y, Y + X - m\}$. Then if the person chooses option $Y + X - m$ even though option $Y$ was available, it means that they were willing to give up an amount of money $m$ in order to change their outcome by $X$. Thus the person is willing to pay the price $m$ to buy $X$ on top of $Y$. We then define the highest such price $m$ as the \emph{buying price} $P_B(X|Y) \equiv \sup \{ m: \exists S \text{ with } \{ Y\} \cap S = \varnothing \text{ s.t. } v(Y + X - m | S_m) \geq \max_{Z \in S \cup \{Y\}} v(Z|S_m) \}$. Similarly, if the person chooses option $Y$ even though option $Y + X - m$ was available, it means that they were willing to give up getting $X$ on top of $Y$ in order to receive the price $m$ --- to receive an extra amount of money $m$. In other words, they are willing to sell $X$ on top of $Y$ for the price $m$. We define the lowest such price $m$ as the \emph{selling price} $P_S(X|Y) \equiv \inf \{ m: \exists S \text{ with } \{Y\} \cap S = \varnothing \text{ s.t. } v(Y|S_m) \geq \max_{Z \in S_m}\}$. Note that the set of buying and selling prices for $X$ on top of $Y$ is never empty: when choosing only between $X$ and $Y$, by non-satiation the person always chooses $X - m$ for sufficiently low $m$, and never chooses $X - m$ --- and hence chooses $Y$ --- for sufficiently large $m$.

Let us now define $\Delta$, the largest price differential between the buying and the selling price for any good: $\Delta \equiv \sup_{X, Y, Y'} P_B(X|Y) - P_S(X|Y')$.\footnote{$\Delta$ could be infinite.} Our first result shows that when the buying price of a good ever strictly exceeds its selling price, then a narrow bracketer can be offered choices such that they throw away an amount of money arbitrarily close to price difference. All proofs are in the appendix.

\begin{proposition} \label{prop:arbitrage}
 If $\Delta > 0$, then for every $\delta \in [0, \Delta)$ there exist choice sets $S_1$ and $S_2$ s.t. a narrow bracketer ends up with a final outcome $A - \delta$ even though $A \in S_1 + S_2$.
\end{proposition}

The intuition for the result is straightforward: if the person is willing to buy $X$ for a strictly larger price on top of $Y$ than she is willing to sell $X$ for when it is on top of $Y'$, then a narrow bracketer facing these two simultaneous choices will buy $X$ for the high price in the first choice, sell it for the low price in the second choice, and thus make a loss equal their difference.

So when $\Delta > 0$, there are choices where a narrow bracketer makes a dominated choice throwing away an amount of money up to size $\Delta$. Assuming monotonicity in money, a broad bracketer would never choose that way, hence offering a person these choices allows us to identify whether the person is a narrow or a broad bracketer.

The next proposition shows that when $\Delta = 0$, so that every bundle $X$ has a unique price, then this price function is an additive utility representation for the choices of this person. 

\begin{proposition}\label{prop:price-representation}
 If $\Delta = 0$, then choices satisfy WARP and there is a utility representation $P(X)$ for the choices with $P(X) = P_B(X|0)$, and this utility representation is additive, satisfying $P(X + Y) = P(X) + P(Y)$.
\end{proposition}

Intuitively, if across all possible choices, the person is willing to buy $X$ for $m$ or less, and sells for $m$ or more, then the value of $X$ across all choice settings is equal to $m$. In particular, this is the price of $X$ when the alternative is $0$. Moreover, since the value of $X$ is in some sense independent of the choice where it is bought or sold, the price of $X + Y$ is the price of $X$ plus the price of $Y$. 

It is easy to see that a person who maximizes an additive utility function such as $P(\cdot)$ will choose the same way whether they are bracketing narrowly or broadly, since $\max_{X, Y} P(X + Y) = \max_{X, Y} P(X) + P(Y) = \max_{X} P(X) + \max_{Y} P(Y)$. Therefore in this case we cannot identify whether the person is a narrow or a broad bracketer. Hence, when $\Delta = 0$, we cannot distinguish between narrow and broad bracketing, while Proposition \ref{prop:arbitrage} shows that when $\Delta > 0$ we can, since a narrow bracketer can be arbitraged. This proves the following result:

\begin{proposition}\label{prop:identification}
    We can distinguish between narrow and broad bracketing --- narrow and broad bracketing are identified --- if and only if $\bar{\Delta} > 0$.
\end{proposition}

By Proposition \ref{prop:price-representation}, we know that $\Delta > 0$ requires that choices violate either WARP or the additivity of the utility representation $P(\cdot)$. We now illustrate how to look for potential price differentials, since these guarantee that we can identify bracketing. 

\subsection{Discussion}

Propositions \ref{prop:arbitrage}, \ref{prop:price-representation}, and \ref{prop:identification} together show that the central error committed by a narrow bracketer is that they neglect price differentials, and thus arbitrage opportunities, across choices. These prices for any bundle $X$ are measured in the amount of some numeraire good --- "money" --- they are willing to give up to obtain the additional bundle $X$. In other words, it is the rate of substitution between $X$ and the numeraire good. Thus, as long as the rate of substitution is constant for any pair of goods across all choices, there are no price differentials and hence the narrow bracketer behaves like a broad bracketer (Proposition \ref{prop:price-representation}). When it varies, then a narrow bracketer fails to arbitrage across choices, and makes choices such that they leave money on the table (Proposition \ref{prop:arbitrage}), which, under the assumption that people never leave money on the table in single choices, implies that we can distinguish between narrow and broad bracketing (Proposition \ref{prop:identification}). 

One consequence of this is that when a person's observed choices are consistent with an additive utility representation as in Proposition \ref{prop:price-representation}, then we cannot identify from those choices whether the person is bracketing narrowly or broadly. Thus when we find that one group of people is less consistent with broad bracketing than another, we cannot infer that this group is more likely to narrowly bracket - an alternative explanation is that their choices are consistent with both. 

Another point worth highlighting is that "money" is defined by its properties in Subsection \ref{subsec:framework-setup-and-definitions}. This means that any good that satisfies the necessary monotonicity properties can serve as "money". For example, when considering repeated choices about how to spend one's time, an increase in the time spent with others rather than alone as well as the probability of achieving ones life goal may both serve as numeraires. Whether this is the case or not may differ from one person to another, since whether something is a good satisfying the monotonicity properties we require depends on preferences (or the assumptions on preferences we are willing to make). With such numeraires, narrow bracketing might lead people to end up in situations where they are strictly worse off, spending more time alone or lowering their likelihood of achieving a life goal more than was possible.

\paragraph{Three sources of price differentials} As we show in Proposition \ref{prop:additivity-conditions} in Appendix \ref{appendix:additive-utility}, additivity of $P(\cdot)$ implies (among other things) that the choices satisfy expected utility and are linear in the amount of work done. Here we illustrate how violating one of these or WARP leads to price differentials. In our experiments, we use these price differentials as a basis to identify whether people bracket these price differentials narrowly or broadly. 

First, consider a person who has a disutility from work equal to $d(e) = 10 \cdot e^2 / 2$ and whose total utility from working $e$ hours and earning \$$m$ is $m - d(e)$. Then this person chooses to do 1 hour of work over 0 hours of work if they are paid at least \$$5$ more, and this person chooses to do 2 hours of work over 1 hour of work if they are paid at least \$$15$ more. Thus, the price differential across these two situations is $15 - 5 = 10$. 

Now consider a person who faces choices over lotteries: they either have to do no work at all, or they have to work for $2$ hours. 
Suppose that this person chooses to receive \$$10$ for sure for doing $2$ hours when a die rolls a 1 or a 2 rather than doing no work for no money, and chooses no money for sure and doing $2$ hours when a die rolls 3 or 4 over \$$5$ for sure for doing $2$ hours of work when a die rolls 3, 4, 5, or 6.
 
Thus they choose do take on a 1/3 increase in probability of doing $2$ hours of work for \$$10$ in the first case, but reject it for \$$5$ in the second, and thus the price differential is \$$5$.

Finally, consider a person who violates WARP as follows: when choosing between doing $2$ hours of work for \$$19$ and doing no work, they choose no work, as would be in line with the disutility above. However, when choosing between $2$ hours of work for \$$19$, doing no work, and doing $2$ hours of work for \$$10$, they choose to do $2$ hours of work for \$$19$. Moreover, suppose they continue choosing $2$ hours of work for as little as \$$15$. Then the price differential is at least \$$4 = 19 - 15$. 

Hence, in each of these cases, we know by Proposition \ref{prop:arbitrage} that we can use these price differentials to construct choices to test bracketing via arbitrage.

\paragraph{Source of price differentials in the Literature}\label{par:source-of-price-differentials} We now review a small subset of the literature that tests bracketing and for each identify potential price differentials underlying their identification.

\cite{tverskyKahneman1981framing} identify choices in which people's choices are consistent with risk-aversion in gains and risk-seeking in losses, so that they reject a given lottery when it changes the gains, yet they accept it in the loss domain. In our terminology, they sell the lottery when it is offered in the gains domain and they buy it in the loss domain. As \cite{rabin2009narrow} point out, it is sufficient if there is a \emph{change} in the degree of risk aversion, which leads to a price differential because this changes their willingness to buy the lottery.

\cite{vorjohann2020referenceDependentChoiceBracketing} defines narrow bracketing as correlation neglect. To give a stark example, consider a person who may not like risky assets, but faces two assets, one of which always goes down when the other goes up. In such a situation, a person who can only buy a single asset might be willing to pay only a low price due to the potential of losses, yet when they can buy both and hedge against losses, their willingness to pay rises. Thus, whenever correlation affects prices, narrow bracketing implies correlation neglect, although it is not equivalent to it in more general settings.

\cite{ellisFreeman2020revealingBracketing} study choices across multiple domains, including portfolio choices with similar sources of price differentials as \cite{tverskyKahneman1981framing} and \cite{rabin2009narrow}. Additionally, they also test bracketing over consumer problems with induced values that are non-linear; as well as over divisions of money between other (anonymous) subjects. In the latter case, if a person is inequity averse, then they may choose an equal split of $(5, 5)$ over a split of $(7, 4)$, as well as over a split of $(4.5, 6)$. Therefore, the price of \$$1$ to person 2 equals \$$2$ from person 1 in the first choice, and \$$0.5$ in the second, leading to price differentials. Note that a person who displays efficiency motives, such as maximizing the total amount of money that the two subjects receive, would maximize an additive function. With such preferences, we would not be able to test whether this person is bracketing narrowly or broadly.

\paragraph{Other sources of price variation} Our result that bracketing is unidentified if and only if there exists an additive utility representation denoted $P(\cdot)$ (for \emph{price}) generalizes to richer choice settings. For instance, suppose that a person chooses how much time to spend on each of two activities, and is (exogenously) required to choose the total amount for activity 1 and separately the total amount for activity 2. If we assume that their utility from working $w_1$ hours on the first and $w_2$ hours on the second activity is $u(w_1, w_2) - (w_1 + w_2)$,\footnote{For simplicity we assume in this example that all remaining time can be used in a linear way for leisure. This can be generalized.} then we can show that bracketing is unidentified if there is some utility $P(\cdot)$ that is additive \emph{over the set of choices we could possibly observe}. This means that $P(X + Y) = P(X) + P(Y)$ has to hold for all $X$ which affect only time spent on activity 1 but not activity 2, and all $Y$ that affect activity $2$ but not 1. It is clear that any additively separable utility function $u(w_1, w_2) = u_1(w_1) + u_2(w_2)$ satisfies this condition, where the $u_i$ need not be linear; and that if it is not additively separable, then a narrow bracketer behaves differently in some choices.\footnote{Given that we do not allow more than 1 choice per activity, we cannot arbitrage, since this would require one choice where we sell and one choice where we buy the same good.} This highlights both the key role of additivity, as well as the necessary variation in order to be able to identify bracketing.

%% file: experiment-1-design-resubmission.tex
In the following sections, we describe the first pre-registered online experiments, written in Lioness \citep{giamattei2020lioness}. Study 1 (conducted between December 2019 and January 2020) tests the effects of narrow and broad bracketing on the willingness to work in a real effort choice setting.\footnote{See https://doi.org/10.1257/rct.3412-4.499999999999999, in particular the 'December Design' under \emph{Supporting Documents and Materials}.} The study consists of four parts. We start by describing the real effort task and the main parts and treatments of Study 1 (detailed instructions are provided in Appendix~\ref{appendix:instructions}).



\subsection{Study 1: Design }

\begin{figure}[H]
    \centering
\includegraphics[scale=0.8]{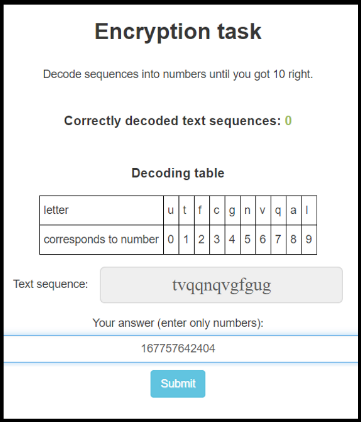}
    \caption{Decoding task}
    \label{fig:task}
\end{figure}

\paragraph{Part 1 Tutorial} The study starts with a tutorial phase where participants begin by familiarizing themselves with the real-effort task through practice trials, continuing until they correctly complete three tasks. Each task involves decoding a sequence of twelve letters into their corresponding numerical values, as illustrated in Figure \ref{fig:task}.

For each sequence, participants receive a newly generated table that randomly assigns ten letters to the digits 0 through 9. This mapping is re-created after each attempt, regardless of whether the previous response was correct. By continuously changing letter-to-number assignments, we introduce a greater learning challenge compared to conventional encryption or typing tasks \citep[e.g.,][]{erkal2011relative, de2017bonus, de2018your}. This design increases the convexity of the effort costs, a key mechanism for satisfying our identifiability assumptions.

\paragraph{Part 2: Elicit Tediousness} After completing the tutorial, participants rate the perceived tediousness of the task on a scale from 1 ("not tedious at all") to 10 ("extremely tedious"). This provides a control variable that is consistent across all treatment conditions.

\paragraph{Part 3: Elicit Reservation Wages} We elicit participants' reservation wage for a high-work option that requires completing 15 additional tasks compared to the low-work option, with the specific conditions varying by treatment. The four treatments (detailed below) differ in how participants’ initial endowment of tasks and money is structured and presented.

Participants make decisions in two sequential choice scenarios, Scenario 1 and Scenario 2. In each scenario, we elicit a reservation wage using an incentivized price list, where participants decide whether to accept or reject the additional workload for a series of incremental \emph{extra} wages ranging from \$0.25 to \$4.00 in increments of \$0.25. These extra wages are offered on top of the \$4.00 baseline payment for the alternative workload.\footnote{Details on individual choices for each treatment are provided in Appendix \ref{appendix:instructions}.}

The framing of payments varies by treatment (see description of Main Treatments). Participants are informed that one binding decision from either scenario will be randomly selected and implemented.

\paragraph{Part 4: Complete Tasks} We randomly determine the binding choice and inform subjects about the total payment and number of sequences to decode. Subjects can complete the tasks without time constraints. We then ask for a short demographic questionnaire and display a summary of total earnings.

\subsection{Treatments}

In our treatments, participants made a single active choice in each scenario. The choice combined an endowment of work and/or money. This structure, often referred to as \emph{endowment bracketing},\footnote{See (\cite{koch2019correlates}, \cite{ellisFreeman2020revealingBracketing}).} is most studied in the context of background risk, where individuals fail to integrate it with other risky choices. In this regard, \cite{barberis2006individual} and \cite{mu2021background} highlight that ignoring background risk helps explain risk aversion over small stakes.\footnote{Failure to fully incorporate endowments can lead individuals to treat money and goods as less fungible than they are, linking endowment bracketing to mental budgeting \citep{heathSoll1996mentalBudgeting, hastingsShapiro2013fungibility, abeler2017fungibility}. Furthermore, as shown in \cite{imas2016realizationEffect}, the behavior of individuals differs between real and paper losses, suggesting that factors beyond bracketing influence the role of endowments in decision-making.}

We report three between-subjects treatments:\footnote{We have several additional treatments that were relevant for our earlier paper titled "Narrow Bracketing in Work Choices" and which we deemphasize in the interest of space after adding study 2. These treatments are MONEY, BEFORE, and AFTER. MONEY is described in more detail in Appendix \ref{app:bracketing-work-and-money}; BEFORE and AFTER are described in Appendix \ref{appendix:before_after_summary}.} NONE, where there is no endowment; MONEY/LOW, where the endowment consists of \$2.00; and BOTH, where the endowment consists of \$2.00 and 15 sequences to decode. NONE and BOTH have identical total choice sets: the total amount of money and work possible is the same for these treatments. The choice of additional work and money is identical in BOTH and MONEY/LOW, but since MONEY/LOW has no endowment of work, each option in MONEY/LOW leads to exactly 15 fewer sequences than in BOTH -- hence MONEY/LOW is the only treatment with different total outcomes.

We show below the text displayed to participants for a single choice of the price list for Scenario 1 for each treatment, while we refer to Appendix \ref{appendix:instructions} for the screens of the choice list items. Scenario-2 choices require exactly 15 sequences more than the Scenario-1 choices of the same treatment. Participants have to choose between a list of Option A and Option B choices, with payments for Option B (in addition to the completion fee) going from $\$4.25$ to $\$8$ in steps of $\$0.25$.

\include{appendix-screens-study1}

\subsection{Bracketing Hypotheses}\label{subsec:hypotheses}

Note that the treatments NONE and BOTH have \emph{identical} total outcomes in the same scenario, but with 15 sequences and $\$2.00$ shifted to the endowment for BOTH. Hence broad bracketing predicts identical choices over \emph{total} outcomes across them.\footnote{In a similar design in online auctions on eBay, \cite{hossain2006plus} find higher revenues and number of bidders when the starting price is reduced by the same amount that the shipping costs are increased. This is consistent with participants ignoring shipping costs (partially), similar to ignoring endowments in our setting.} MONEY/LOW on the other hand has identical Options in the choice list as BOTH, even though it has no work endowment and hence require 15 fewer tasks. So the choice options \emph{ignoring} endowments are identical for MONEY/LOW and BOTH, so that narrow bracketing predicts identical choices of options -- but not of total outcomes. Denoting by $m_T$ the average reservation price elicited in treatment $T$, we get the following hypotheses:\footnote{One can formally relate these hypotheses to our theoretical framework by noting that situations of a single choice with a given endowment correspond to situations where the second choice set is a singleton: $\mathcal{Y} = \{Y\})$.}

\begin{hypothesis}[Broad Bracketing]\label{hyp:broad}
  Behavior is consistent with broad bracketing if $m_{NONE} = m_{BOTH}$ in every Scenario.
\end{hypothesis}

\begin{hypothesis}[Narrow Bracketing]\label{hyp:narrow}
  Behavior is consistent with narrow bracketing if $m_{BOTH} = m_{MONEY/LOW}$ in every Scenario.
\end{hypothesis}

As we discussed in Section \ref{sec:framework}, we can cannot always identify bracketing (that is, distinguish narrow from broad bracketing). In the case of our experiment, we need that $m_{MONEY/LOW} \neq m_{NONE}$: if $m_{MONEY/LOW} = m_{NONE}$, then choices are consistent with linear preferences, so that we will either fail to reject both narrow and broad bracketing or reject both simultaneously. This will happen if participants are as willing to do 15 additional sequences on top of 0 sequences, as they are willing to do them on top of 15 sequences.

\begin{assumption}[Identification Assumption]\label{assumption:ID}
  We can identify narrow vs broad bracketing if and only if $m_{NONE} \neq m_{MONEY/LOW}$.
\end{assumption}

Since we have two Scenarios, we have two tests for broad and two tests for narrow bracketing. For tests with $5\%$ significance, we would therefore apply the Bonferroni correction of rejecting the null hypothesis in a given Scenario only if $m_{NONE}$ differs from $m_{BOTH}$ ($H_{0}:$ broad bracketing) or from $m_{MONEY/LOW}$ ($H_{0}:$ narrow bracketing) at the $2.5\%$-level, to avoid overrejection based on having two tests.

\subsection{Study 1: Summary Statistics}

We recruited in total $716$ subjects on Amazon Mechanical Turk between the end of December 2019 and the beginning of February 2020. Table \ref{tab:session_summary} in Appendix \ref{appendix:additional-results} shows how many participants and in which treatment we recruited by session. In Table \ref{tab:summary_statistics} we report summary statistics across treatments. Of the subjects recruited, 127 did not complete the experiment (17.7\% attrition rate). Although differential attrition across treatments could itself hint at the consequences of narrow bracketing, we see no evidence for it (see Table \ref{tab:attrition} in Appendix \ref{appendix:additional-results}). Across all treatments, between half and two-thirds of participants failing to complete the HIT drop before completing the practice tasks, while the remaining others drop out after finding out how many tasks they have to do in total. Treatments are similar in terms of gender composition ($\chi^2$ test p-value: $0.84$), while participants are slightly older in the NONE treatment compared to other treatments ($37.8$ years vs $35.0$-$35.1$, $\chi^2$ test p-value: $0.29$). Finally, individuals rate the task on average as $7.33$-$7.54$ out of $10$ in tediousness, which does not significantly vary across treatments ($\chi^2$ test p-value: $0.86$). Roughly $18\%$ of the choices made within a choice-list are inconsistent: in a few cases subjects make only one inconsistent choice, while in other cases choices are inconsistent throughout the list of wages offered (such as when they switch repeatedly between the options, even though we monotonically increase payment). In our main analysis we drop a scenario if individuals make more than one inconsistent choices in it. To detect an effect size of 0.40 at a 5\% level of significance with 90\% power, we would need 174 observations in BOTH and 116 in MONEY/LOW and NONE treatments. The number of observations collected with consistent choices are above these thresholds and therefore considered sufficient for our treatments comparisons, although as the discussion on identification makes clear, the effect size decreases as preferences become more linear. Participants earned $\$7.30$ on average, for an average working time of 35 minutes.\footnote{Subjects' feedback rated on average this payment as generous. For details, see \href{https://turkerview.com/requesters/A3TEY5GKYRHXWG}{https://turkerview.com/requesters/A3TEY5GKYRHXWG}.}

\import{./data/}{summary_statistics_main.tex}

%% file: appendix-screens-study1.tex
\newpage

\paragraph{NONE: No Endowment}\mbox{}\\

\begin{tcolorbox}
\textbf{By completing the HIT you will receive a total payment (which includes the \$2.00 completion fee) depending on your choices.}\\

\vspace{-1cm}
\begin{center}\small{
\flushleft{\begin{tabular}{l l c c l}\hline
&& OPTION A & OPTION B & \\ \hline
5) & {\raggedright 15 sequences for a total payment of \$6.00} & $\bigcirc$ & $\bigcirc$ & {\raggedright 30 sequences for a total payment of \$7.25} \\
\hline
\end{tabular}}}
\end{center}
\end{tcolorbox}

\paragraph{BOTH: Both Money and Work Endowment}\mbox{}\\

\begin{tcolorbox}
\textbf{Note: you are required to decode 15 sequences correctly, in addition to the sequences based on your choices.}\\
\textbf{By completing the HIT you will receive \$2.00 plus a bonus depending on your choices.}\\

\vspace{-1cm}
\begin{center}\small{
\flushleft{\begin{tabular}{l l c c l}\hline
&& OPTION A & OPTION B & \\ \hline
5) & {\raggedright 0 additional sequences for an extra \$4.00} & $\bigcirc$ & $\bigcirc$ & {\raggedright 15 additional sequences for an extra \$5.25} \\
\hline
\end{tabular}}}
\end{center}
\end{tcolorbox}

\paragraph{MONEY/LOW: Money but no Work Endowment}\mbox{}\\

\begin{tcolorbox}
\textbf{By completing the HIT you will receive \$2.00 plus a bonus depending on your choices.}\\

\vspace{-1cm}
\begin{center}\small{
\flushleft{\begin{tabular}{l l c c l}\hline
&& OPTION A & OPTION B & \\ \hline
5) & {\raggedright 0 sequences for an extra of \$4.00} & $\bigcirc$ & $\bigcirc$ & {\raggedright 15 sequences for an extra of \$5.25} \\
\hline
\end{tabular}}}
\end{center}
\end{tcolorbox}

%% file: data/summary_statistics_main.tex
\begin{table}[!h]
\centering
\caption{\label{tab:summary_statistics}Summary statistics for main treatments.}
\centering
\fontsize{12}{14}\selectfont
\begin{tabular}[t]{lrrrr}
\toprule
\textbf{} & \textbf{NONE} & \textbf{BOTH} & \textbf{MONEY/LOW} & \textbf{p-value}\\
\midrule\\
Participants & 200 & 320 & 196 & \\
Attrition & 18\% & 20.3\% & 13.3\% & 0.13\\
Final Participants & 164 & 255 & 170 & \\
\midrule\\
Share Female & 0.4 & 0.38 & 0.4 & 0.84\\
Age & 37.8 & 35 & 35.1 & 0.29\\
Tediousness & 7.54 & 7.45 & 7.33 & 0.86\\
\midrule\\
\addlinespace[0.3em]
\multicolumn{5}{l}{\textbf{Inconsistent Choices}}\\
\hspace{1em}Scenario 1 & 17\% & 18.8\% & 20.4\% & 0.69\\
\hspace{1em}Scenario 2 & 15\% & 18.4\% & 18.9\% & 0.52\\
\bottomrule
\end{tabular}
\end{table}

%% file: data/experiment-results.tex
We now analyse the reservation wages across treatments: the smallest extra wage for which subjects prefer OPTION B over OPTION A, where OPTION B always requires decoding correctly 15 sequences more than OPTION A. The extra payments start at $\$0.25$, if a subject always accepts the extra work, and increases in $\$0.25$ increments to $\$4.00$. If a subject never accepts the extra work, we code the reservation wage as $\$4.25$.

\subsection{Main Results}

\begin{figure}[H]
  \centering
  \includegraphics[scale=0.8]{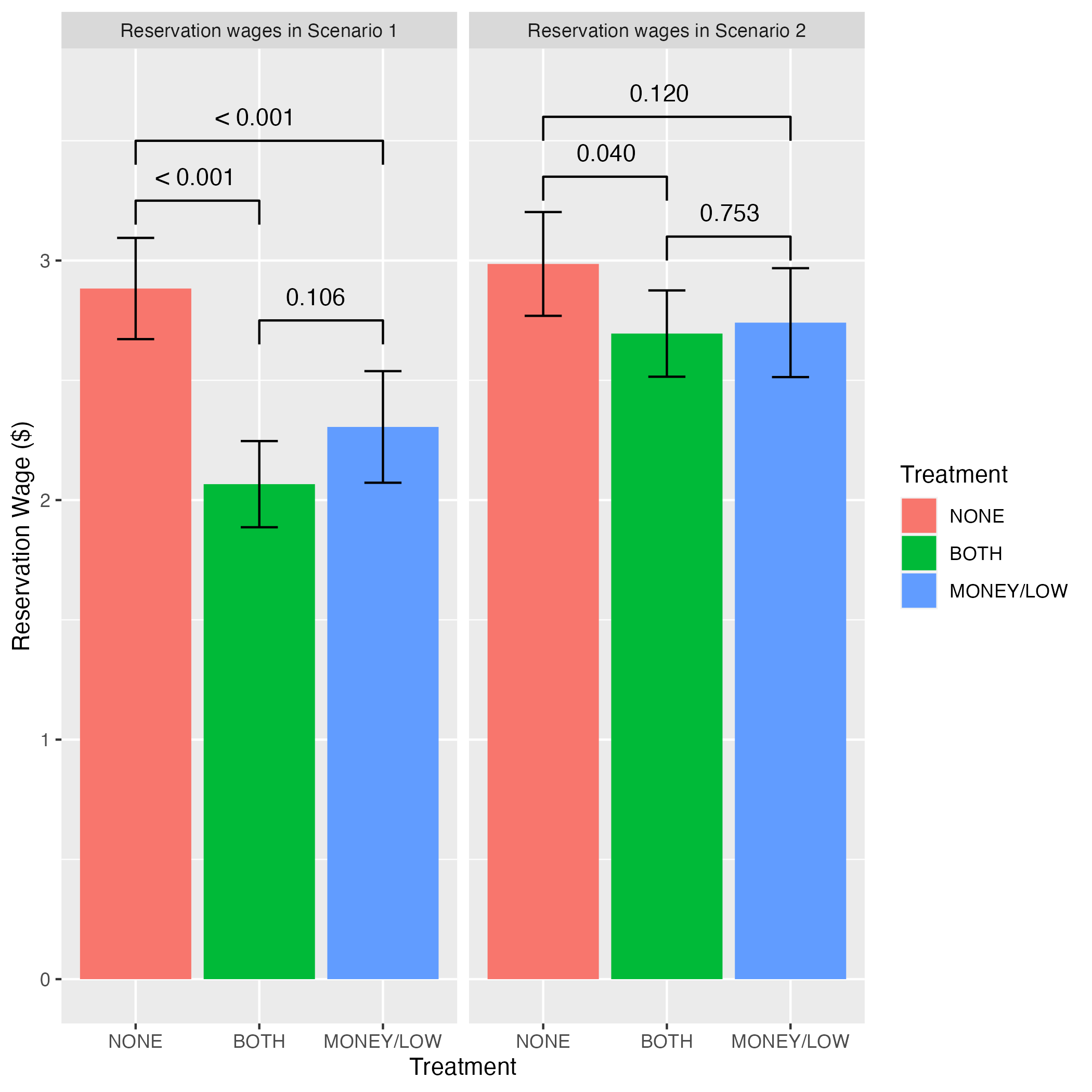}
    \caption{The reservation wages by treatment for each of the two scenarios, along with confidence intervals (2 standard errors above and below the estimate). The p-values compare average reservation wages between two treatments via two-sided t-tests.}
    \label{fig:bar-plot-means}
\end{figure}

\begin{result}
We reject Hypothesis \ref{hyp:broad} that individuals bracket decisions broadly.
\end{result}

Based on figure \ref{fig:bar-plot-means}, we reject broad bracketing as per Hypothesis \ref{hyp:broad}, since we reject broad bracketing in one of the scenarios (Scenario 1) at the Bonferroni-corrected p-value of less than $2.5\%$-significance (see our discussion in \ref{subsec:hypotheses}).\footnote{See \ref{raw_data_plots} in Appendix \ref{appendix:additional-results} for bar plots and kernel density plots of the raw reservation wage data by treatment and scenario.} Concretely, the average extra reservation wage in treatment NONE in Scenario 1 is $\$2.88$ compared to $\$2.07$ in BOTH, and this difference is significant at the $0.001$-level. Note that identifying assumption \ref{assumption:ID} holds in Scenario 1, since the reservation wage in NONE and in MONEY/LOW are not equal.\footnote{Since we keep the choice sets exactly equal across treatments, we ensure that differences in behavior are not because of (non-bracketing) interactions with work they do outside of our experiment. For example, if participants could earn less in some treatments, they might decide to spend less time on our experiment and work more on other tasks on MTurk instead. If our participants did that, it would by itself be a sign that they failed to realize that their overall outcomes are actually the same.}

Notice that the Scenario-2 reservation wage of $\$2.74$ in MONEY/LOW and of $\$2.88$ in NONE are not statistically significantly different (p-value of $0.120$). Thus identifying assumption \ref{assumption:ID} fails in this Scenario, and consequently we fail to reject broad bracketing in this scenario. We also fail to reject narrow bracketing in this scenario: the reservation wage of $\$2.70$ for BOTH is not statistically significantly different from MONEY/LOW. Scenario 2 lacks power to identify bracketing, because we cannot rule out constant marginal disutility at 15 and 30 baseline sequences.

Since the difference in Scenario 1 between the reservation wage of $\$2.31$ in MONEY/LOW and of $\$2.07$ in BOTH is not significant (p-value: $0.106$), we cannot reject narrow bracketing in either Scenario and hence do not reject it overall. Since the identification assumption holds in Scenario 1, the identification assumption holds, so the failure to reject it does not stem from a failure of the identification assumption to hold.

\begin{result}
We fail to reject Hypothesis \ref{hyp:narrow} that individuals bracket decisions narrowly.
\end{result}

We perform similar comparisons between the treatments via Wilcoxon rank-sum tests as well as when restricting to the balanced sessions only, both of which lead to identical conclusions -- see Appendix \ref{appendix:additional-results} for details.\footnote{In addition to using the Wilcoxon rank-sum tests, we also perform tests for different sessions of BOTH. Specifically, in the initial sessions, we mistakenly displayed the endowments on the page right before the first choice page. We fixed this, displaying it \emph{on} the first choice page only, which is why we collected more data for treatment BOTH. See \ref{appendix:reveal-on-page-only} and \ref{appendix:reveal-before-page} for the same results when we restrict treatment BOTH to when we display the endowments on the first choice page only, or when we display it right before the first choice page. See \ref{appendix:balanced} for when we restrict the treatments to those sessions in which data collection was balanced -- since some sessions were not balanced. In all these cases, we reject broad bracketing in Scenario 1 at the $2.5\%$-level and hence overall, and do not reject narrow bracketing in either Scenario 1 or Scenario 2.}

Overall, our results can be summarized by saying that participants average behavior is consistent with a disutility function $d(\cdot)$ satisfying $d(15) \approx 2.3$ (based on Scenario 1, MONEY/LOW), with $d(30) - d(15) \approx 2.8$ (based on Scenario 1, NONE; and Scenario 2, MONEY/LOW) and with $d(45) - d(30) \approx 3.00$ (based on Scenario 2, NONE); and that participants narrowly bracket these preferences. The disutility of effort is thus convex, and growing less convex with additional tasks.\footnote{We report in Appendix \ref{appendix:before-after} the description and results of a follow-up study where we attempt to debias subjects. However, the results confirm the effect of narrow bracketing in work choices.}

%% file: experiment-2-design-resubmission.tex
Following our theory, Study 2 aims to test for narrow and broad bracketing by generating variation in prices in three different domains in a within-subjects design.\footnote{The design of Study 2 was preregistered in OSF: \href{https://osf.io/5q7wn/}{https://osf.io/5q7wn/}.}

\subsection{Design}

Study 2 follows the structure of Study 1, with the main differences occurring in Part 3 where we elicit willingness to work. After assessing perceived tediousness, subjects received instructions on the choices to be made and answered two comprehension multiple choice questions.\footnote{The questions are reported in Appendix \ref{appendix:study2-instructions}. These questions were crucial to understanding the experimental incentives.} Only those who correctly answered advanced to the decision stage, where they responded to 35 scenarios that involved different work options at varying pay levels.\footnote{See Appendix \ref{appendix:study2-instructions} for the instructions of the main task as well as the individual Scenarios in study 2.} Following our theoretical framework, these scenarios were aimed at identifying participants in three domains that we label 'Linear', 'WARP', and 'Probability', where we attempted to find price variation by finding violations of linear utility, of WARP, and of expected utility respectively. 

We now describe how we attempted for each domain attempt to find a pair of choices with a buying price in one choice that is higher than the selling price in another choice, so as to use these choices to arbitrage the participant. 

\paragraph{Linear} Identification in the linear domain was performed in two steps. First, we elicited the maximum number of sequences participants were willing to decode (up to 25) at five different piece rates, ranging from £0.02 to £0.12 per sequence. The participants then responded to follow-up questions adapted from prior responses, presenting choices between their previous selection and an alternative involving five more (or fewer) sequences at the same piece rate. Example:

    \begin{table}[H]
\begin{tabular}{p{7cm}|p{0.1cm}|p{7cm}}
\textbf{Scenario 1: Choose your preferred option:} &  & \textbf{Scenario 2: Choose your preferred option:} \\ $\bullet$ Option A: Decode 10 sequences for £1.20.         &  & $\bullet$ Option C: Decode 10 sequences for £0.80. \\
$\bullet$ Option B: Decode 15 sequences for £1.80.      &  &  $\bullet$ Option D: Decode 5 sequences for £0.40. \\
\end{tabular}
\end{table}

Participants are identified in this domain if we find two scenarios such that they refuse to decode more sequences at a higher price (e.g., choosing Option A in Scenario 1 ond thus refuse 5 tasks for £0.60) and accept to decode more sequences at a lower price (e.g., choosing Option C in Scenario 2 and thus accept 5 tasks for £0.40).

\paragraph{WARP} WARP violations were assessed by presenting choices between tasks with different combinations of effort and payment. In some of the choices (as in Scenario 4) we included a third option (a decoy) to influence the attractiveness of the other item.\footnote{Following \cite{kaptein2016tracking} we tested for different decoys as well as choices presented in different orders.} Example:

\begin{table}[H]
\begin{tabular}{p{7cm}|p{0.1cm}|p{7cm}}
\textbf{Scenario 3: Choose your preferred option:} &  & \textbf{Scenario 4: Choose your preferred option:} \\
$\bullet$ Option A: Decode 5 sequences for £0.45.         &  & $\bullet$  Option C: Decode 5 sequences for £0.45.  \\
$\bullet$ Option B: Decode 12 sequences for £0.80. &  & $\bullet$ 
 Option D: Decode 12 sequences for £0.75.          \\
& & $\bullet$ Option E: Decode 12 sequences for £0.70.
\end{tabular}
\end{table}

Participants were identified in this setting if they selected Option A in Scenario 3 but chose Option D in Scenario 4. These choices are consistent with a WARP violation.

\paragraph{Probability} In the probability domain, we identify subjects presenting choices involving probabilistic work obligations and sure payments. Example:

    \begin{table}[H]
\begin{tabular}{p{7cm}|p{0.1cm}|p{7cm}}
\textbf{Scenario 5: Choose your preferred option:} &  & \textbf{Scenario 6: Choose your preferred option:} \\ $\bullet$ Option A: Receive £0.50 for sure. If the six-sided die
rolls 5 or 6, decode 20 sequences; otherwise, decode 0 sequences. &  & $\bullet$ Option C: Receive £0.70 for sure. If the six-sided die
rolls 1, 2, 3, or 4, decode 20 sequences; otherwise, decode 0 sequences. \\
$\bullet$ Option B: Decode 0 sequences for £0.25. &  &  $\bullet$ Option D: Receive £0.50 for sure. If the six-sided die rolls 3 or 4, decode 20 sequences; otherwise, decode 0 sequences. \\
\end{tabular}
\end{table}

Participants are identified in the probability domain if they choose Option B in Scenario 5 and thus reject receiving an extra doing 20 sequences if a 5 or a 6 is rolled and Option C in Scenario 6, that is, rejecting an increase in the probability of decoding 20 sequences for £0.25, but accepting the same increase in probability for £0.20.

\paragraph{Arbitrage} After the identification phase, subjects completed scenarios that tested for narrow and broad bracketing, following our arbitrage approach. In \emph{Simultaneous} scenarios we displayed the two scenarios on a single screen. In the instructions we specify that when two decisions are shown on the same Scenario page, they are both relevant for the amount of work and payment.\footnote{To ensure that participants understand this instructions, and in particular the way the amount of work and bonuses are defined, we ask them to answer two comprehension questions before proceeding to the decision part of the experiment. Subjects who fail to correctly answer the two questions are excluded from the participation in the experiment. The questions are shown in Appendix \ref{appendix:controlquestions}.} We show below the \textit{Simultaneous} version for Scenarios 1 and 2:\footnote{See Appendix \ref{appendix:screensimultaneous} for a screenshot of the Scenario implemented.}

\begin{table}[H]
\begin{tabular}{p{7cm}}
\textbf{Scenario 7 [SIMULTANEOUS LINEAR]} \\
In this Scenario, you have to make two decisions.\\
\textbf{Decision 1: Choose your preferred option:}\\
$\bullet$ Option A: Decode 10 sequences for £1.20.\\
$\bullet$ Option B: Decode 15 sequences for £1.80.\\
Before answering, read the next decision\\
\textbf{Decision 2: Choose your preferred option:}\\
$\bullet$ Option A: Decode 10 sequences for £0.80.\\
$\bullet$ Option B: Decode 15 sequences for £0.40.\\
\end{tabular}
\end{table}

\emph{Combined} scenarios combined them into a single choice. We show below an example of the combined linear case (equivalent to the simultaneous Scenarios 1 and 2), combined probability (equivalent to the simultaneous Scenarios 5 and 6) and combined WARP (equivalent to the simultaneous Scenarios 3 and 4):

\begin{table}[H]
\begin{tabular}{p{6.5cm}|p{0.1cm}|p{9cm}}
\textbf{Scenario 8 [COMBINED LINEAR]: Choose your preferred option:} && \textbf{Scenario 9 [COMBINED PROBABILITY]: Choose your preferred option:} \\ $\bullet$ Option A: Decode 25 sequences for £2.60. && $\bullet$ Option A: Receive £1.20 for sure. If the six-sided die rolls 1, 2, 3, 4, 5 or 6, decode 20 sequences; otherwise, decode 0 sequences.\\
$\bullet$ Option B: Decode 20 sequences for £2.20. && $\bullet$ Option B: Receive £1.00 for sure. If the six-sided die rolls 3, 4, 5 or 6 decode 20 sequences; otherwise, decode 0 sequences. \\
$\bullet$ Option C: Decode 20 sequences for £2.00. && $\bullet$ Option C: Receive £0.95 for sure. If the six-sided die rolls 1, 2, 3 or 4 decode 20 sequences; otherwise, decode 0 sequences.\\
$\bullet$ Option D: Decode 15 sequences for £1.60. && $\bullet$ Option D: Receive £0.75 for sure. If the six-sided die rolls 3 or 4 decode 20 sequences; otherwise, decode 0 sequences.\\
\end{tabular}
\end{table}

\begin{table}[H]
\begin{tabular}{p{6.5cm}p{0.1cm}p{6.5cm}}
\multicolumn{3}{l}{\textbf{Scenario 10 [COMBINED WARP]: Choose your preferred option:}}\\ $\bullet$ Option A: Decode 10 sequences for £0.90.&& $\bullet$ Option D: Decode 17 sequences for £1.25.\\
$\bullet$ Option B: Decode 17 sequences for £1.20. &&$\bullet$ Option E: Decode 24 sequences for £1.55.\\
$\bullet$ Option C: Decode 17 sequences for £1.15.&&
$\bullet$ Option F: Decode 24 sequences for £1.50.\\
\end{tabular}
\end{table}

The remaining part of study 2 is similar to study 1, with one exception. We have included two questions throughout the study to check for attention.\footnote{\citep[see, e.g.,][]{fallucchi2022coordinating}} We use the answers to these questions as a control of attentiveness. 

\subsection{Categorization}\label{subsec:hypotheses}

We do not anticipate that every participant will be identified in one or more domains. To discern which subjects can be identified within each domain, we adhere to Proposition \ref{prop:identification}. Conditional on subjects being identified, they will be classified into one of the following three mutually exclusive categories:\\

\textbf{Narrow Bracketers.} Behavior is consistent with narrow bracketing if the decisions in the 'Narrow' scenario are equivalent to the two equivalent choices when they are presented individually.\\

\textbf{Broad Bracketers.} Behavior is consistent with broad bracketing if the overall amount of work and payment in the 'Narrow' Scenario are equal to those in the 'Broad' Scenario.\\

\textbf{Unknown.} Behavior is not consistent with either narrow and broad bracketing if the decisions in the simultaneous (but separate) Scenario differ from the two choices when made individually (rejecting narrow), and when the decisions in the simultaneous Scenario differ from the decision in the combined-choice Scenario (rejecting broad). 

While our primary interest is on subjects that are ex-ante identified, those who are unidentified will also answer a set of questions with 'simultaneous' and 'combined' Scenarios. In principle, these subjects can also be classified under the three aforementioned categories. However, unlike the identified subjects, it is likely that many of them will be unidentifiable, and hence be categorized both as narrow and broad bracketers at the same time. 

\subsection{Study 2: Summary statistics}\label{subsec:stats2}
We recruited 572 subjects on Academic Prolific in January 2025. Of these, 59 failed to answer the comprehension questions and were excluded from participation. 
The remaining 513 subjects comprised 52.05\% females and 46.59\% males, while 1.36\% are non-binary or have not disclosed their gender. Most participants were aged 18-34, with less than 17\% over 45. On average, participants rated the task as 6.59 out of 10 in tediousness,\footnote{The software failed to record the self-reported tediousness of two subjects.} earned on average £3.79, and decoded on average 11.09 sequences, with an average study time reported in Prolific of approximately 28 minutes.\footnote{The payment aligned with the platform standards, exceeding £6 per hour.}

%% file: experiment-2-results.tex
We begin by examining the number of participants identified in our experiment as shown in Table \ref{tab:summary_statistics2}. Out of 513 subjects, and in each of the three domains approximately 30\% are identified. Examining the distribution of the identified subjects, we observe that 39\% of the sample is not identified in any domain, 34.52\% are identified in a single domain, 22.82\% in two domains, and merely 3.57\% in all three domains.\footnote{We examined correlations across domains and found the highest correlation between WARP and Probability at 17\%.}

\import{./}{data/summary-study2.tex}

\begin{result}\label{res:arbitrage-categorization}
    The majority (50.30\%) of identified subjects are classified as narrow bracketers in LINEAR. The share of broad bracketers is higher than the share of narrow bracketers in WARP (38.16\% vs 26.32\%) and PROBABILITY (46.26\% vs 13.61\%). 
\end{result}

Table \ref{tab:identification-study2} shows the classification of individuals according to their decision-making patterns. A participant is categorized as a narrow bracketer when their two working choices under the 'Simultaneous' choice are consistent with those when considered in isolation. This is observed in 50.30\% of cases for Linear, 26.32\% for WARP,  and 13.61\% for Probability among identified participants. In contrast, a Broad bracketer is someone whose decisions is the same between the 'Simultaneous' and 'Combined' choices, indicating that the total selected workload and payment are unchanged when the two 'Narrow' choices are aggregated. Broad bracketing occurs more frequently in WARP (38.16\%) and Probability (46.26\%) compared to Linear (14.57\%). Overall, of the 305 individuals identified in at least one domain, 43.28\% are classified as narrow bracketers at least once.

We also observe a group of subjects inconsistent with both Narrow and Broad bracketing, which we categorize as 'Unknown' above. These individuals 
account for 35.29\% in WARP, 33.77\% in Linear, and 40.14\% in Probability. 
\import{./}{data/identification-study2.tex}

\paragraph{Exploratory Analysis} Next, we explore the correlates of Narrow bracketing in the three domains. We investigate potential gender differences in the likelihood of being identified as a narrow bracketer, as well as controlling for the perceived tediousness of the task and whether subjects failed the attention checks during the study. The results of these analyses are presented in Table \ref{tab:reg-narrowbracket}.


\import{./}{data/reg_narrow_bracket.tex}

As shown by the regression when we check for gender differences in narrow bracketing, we find contrasting results: in the linear domain, women are significantly less likely to be narrow bracketers than others; in WARP and probability we find an opposite sign, and in probability the likelihood of women to be identified as narrow bracketer is significantly higher.\footnote{$\chi^{2}$ tests confirm similar results: $\chi^{2} (1)=4.318$ $p-value=0.038$ in linear, $\chi^{2} (1)=0.449$ $p-value=0.573$ in WARP and $\chi^{2} (1)=4.463$ $p-value=0.035$ in probability.} In short, there is no robust evidence for systematic gender differences. We also note that perceived tediousness and our measure of attentiveness do not affect the likelihood of being a narrow bracketer in any domain.


\import{./}{data/notidentified.tex}

Unlike most previous studies, our analysis focuses on the subjects identified following our arbitrage condition. We provide the categorization of subjects that are unidentified in table \ref{tab:not-identified}, which shows that the share of subjects that can be categorized as narrow bracketers is much higher, but so is the share of broad bracketers. Notably, at least half of the subjects participating in the study make choices that are consistent with both narrow and broad bracketing in all domains (66\% in Linear, 50\% in WARP and 54\% in Probability). Conditional on not making consistent choices with both narrow and broad bracketing, also for the unidentified subjects we find in Linear the higher share of narrow bracketers; similarly, data confirm the higher share of broad bracketers in Probability; in WARP results differ between identified and unidentified, with a higher share of narrow bracketers in the latter.

%% file: data/summary-study2.tex
\begin{table}[!h]

\caption[Caption]{\label{tab:summary_statistics2}Summary statistics for study 2.\protect\footnotemark}
\centering
\fontsize{12}{14}\selectfont
\begin{tabular}[t]{lrrr}
\toprule
Participants & \multicolumn{3}{c}{572}\\
Failed Comprehension & \multicolumn{3}{c}{59}\\
Final Participants &  \multicolumn{3}{c}{513}\\
\midrule
\textbf{} & \textbf{LINEAR} & \textbf{WARP} & \textbf{PROBABILITY}\\
\midrule
Identified & 149 & 152 & 147 \\
Share female (\%) & 54.36 & 47.37 & 56.46\\
Tediousness & 6.97 & 5.92 & 6.18\\
Share of inattentive (\%) & 8.05 & 9.87 & 12.24 \\
\midrule 
Unidentified & 355 & 361 & 366 \\
Share female (\%) & 51.27 & 54.02 & 50.27\\
Tediousness & 6.41 & 6.87& 6.76\\
Share of inattentive (\%) & 13.80 & 13.30 & 12.30\\
\bottomrule
\end{tabular}
\end{table}

\footnotetext{Nine participants were excluded from the LINEAR case analysis due to software errors displaying incorrect values in the linear narrow scenario. Additionally, the software did not record the self-reported tediousness for two subjects.}

%% file: data/identification-study2.tex
\begin{table}[H]
\caption{\label{tab:identification-study2}Share of narrow and broad bracketers among identified subjects.}
\centering
\fontsize{12}{14}\selectfont
\begin{tabular}[t]{lrrr}
\toprule
\textbf{} & \textbf{LINEAR} & \textbf{WARP} &  \textbf{PROBABILITY} \\
\midrule
\textbf{Identified} & 149 & 152 & 147 \\ \hdashline
Consistent with Narrow Bracketing & 76 (50.33\%) & 40 (26.32\%) &  20 (13.61\%)\\
Consistent with Broad Bracketing & 22 (14.57\%) & 58 (38.16\%) & 68 (46.26\%)\\
Unknown & 51 (33.77\%) & 54 (35.29\%) & 59 (40.14\%)\\
\midrule
\textbf{Unidentified} & 355 & 361 & 366\\
\midrule
Total & 504 & 513 & 513 \\
\bottomrule
\end{tabular}
\end{table}

%% file: data/reg_narrow_bracket.tex
\begin{table}[H]
\caption{\label{tab:reg-narrowbracket} Likelihood of being a narrow bracketer (robust standard errors in parentheses).}
\centering
\fontsize{12}{14}\selectfont
\begin{tabular}[t]{lrrrr}
\toprule
\textbf{Dependent: narrow bracketer} & \textbf{LINEAR} & \textbf{WARP} & \textbf{PROBABILITY} \\
\midrule
\textbf{Female} & $\underset{(0.083)^{\phantom{***}}}{-0.165^{*\phantom{**}}}$ & $\underset{(0.072)^{\phantom{***}}}{0.046^{\phantom{***}}}$ & $\underset{(0.057)^{\phantom{***}}}{{0.147}^{**\phantom{*}}}$\\
\textbf{Tediousness} & $\underset{(0.017)^{\phantom{***}}}{0.006^{\phantom{***}}}$ & $\underset{\phantom{*}(0.014)^{\phantom{**}}}{-0.026^{\phantom{***}}}$ & $\underset{(0.010)}{0.013}^{\phantom{***}}$\\
\textbf{Inattentive} & $\underset{\phantom{*}(0.153)^{\phantom{***}}}{-0.069^{\phantom{***}}}$ & $\underset{(0.132)^{\phantom{***}}}{0.064^{\phantom{**}}}$ & $\underset{(0.093)}{0.042}^{\phantom{**}}$\\
\textbf{Constant} & $\underset{(0.142)^{\phantom{***}}}{0.560^{***}}$ & $\underset{(0.103)^{\phantom{***}}}{0.390^{***}}$ & $\underset{\phantom{***}(0.071)}{-0.034}^{\phantom{***}}$\\
\midrule
\textbf{Total} & 149\phantom{**} & 151\phantom{**} & 146\phantom{**} \\
\hline
\hline \\[-4.5ex] 
\multicolumn{4}{r}{Note: Standard errors in parentheses. $^{*}p \leq 0.05$; $^{**}p \leq 0.01.$; $^{***}p \leq 0.001.$}
\end{tabular}
\end{table}

%% file: data/notidentified.tex
\begin{table}[H]
\caption{\label{tab:not-identified}Narrow and Broad Bracketing among unidentified.}
\centering
\fontsize{12}{14}\selectfont
\begin{tabular}[t]{lrrr}
\toprule
\textbf{} & \textbf{LINEAR} & \textbf{WARP} &  \textbf{PROBABILITY} \\
\midrule
\textbf{Total} & 355 & 361 & 366\\
\midrule
Consistent with Narrow and Broad Bracketing & 233 (66.01\%) & 181 (50.14\%) &  199 (54.37\%)\\
\midrule
\multicolumn{3}{l}{Conditional on not being consistent with Narrow and Broad Bracketing}\\
\midrule
Consistent with Narrow Bracketing & 64 (53.33\%) & 61 (38.89\%) &  38 (22.75\%)\\
Consistent with Broad Bracketing & 16 (13.33\%) & 25 (13.89\%) & 56 (33.33\%)\\
Unknown & 40 (33.33\%) & 94 (52.22\%) & 73 (43.45\%)\\
\bottomrule
\end{tabular}
\end{table}

%% file: conclusion.tex
In this paper, we show that narrow bracketing leads to different behavior from broad bracketing if and only if a person values some bundle differently in different situations, in which case they can be arbitraged across two choices buying the bundle for more in one choice than they sell it for in another. 

We use this type of arbitrage in our main set of experiments to identify bracketing when the source for the change in payment is due to (i) non-linear disutility from work, (ii) violations of expected utility, or (iii) violations of the weak axiom of revealed preference (WARP). We find similar proportions of narrow bracketers as the previous literature for the first source, both in our between-subjects design and our within-subjects design, but the majority of people either brackets broadly or is inconsistent with either narrow or broad bracketing for the latter two sources. One (ex post) hypothesis for this difference is that these types of choices are more easily perceived as connected: in the case of probability choices, we highlight that the outcome of both choices depend on the outcome of the same roll of a die; thus one has to think of this same dice roll when making each choice; in the case of choices with decoys that affect choices, it may be that the mere act of seeing the decoy shifts perceptions for both choices that are displayed and made simultaneously. Varying the source in other ways while testing narrow bracketing may provide a promising avenue for identifying the mechanism behind narrow bracketing, helping us to identify factors of choices that increase or reduce the amount of narrow bracketing. Our paper and experiments are focused on measuring when narrow bracketing occurs, and do not provide direct evidence on the mechanisms.

Finally, the fact that about one third of identified participants are 
not consistent with either narrow or broad bracketing raises further questions. Several natural reasons present themselves: it may be that people are simply confused or inattentive, although the fact that the proportion is similar for choices over simple work-money trade-offs as well as for lotteries suggests otherwise. Alternatively, people may randomize their choices as we know they do even for identical but repeated choices (\cite{agranov_stochastic_2017}), a possibility that we do not account for. Finally, participants may genuinely change their decisions in one choice when facing other choices unlike narrow bracketers, but in a way that differs from how a broad bracketer would do it. Modeling such partial bracketing and disentangling alternative theories requires a different design from our experiments, but will hopefully benefit from the insights and approach outlined in this paper.

%% file: appendix.tex
\import{./}{appendix-proofs.tex}

\import{./}{appendix-instructions.tex}

\import{./}{appendix-additional-results.tex}

\import{./}{appendix-instructions-study2.tex}

\import{./}{appendix-extra-treatments.tex}

%% file: appendix-proofs.tex
\section{Appendix: Proofs}\label{appendix:proofs}

\paragraph{Proof of Proposition \ref{prop:arbitrage}.}

\begin{proof}
    Since $\Delta > 0$, for every $\delta \in [0, \Delta)$, there exists some $X$, $Y$, and $Y'$ s.t. $P_B(X|Y) - P_S(X|Y') > \delta$, hence there exist choice sets $S$ and $S'$ such that $X + Y - m$ is chosen from $S_1 \equiv \{X + Y - m, Y\} \cup S$ and $Y'$ is chosen from $S_2 \equiv \{X + Y' - m', Y'\} \cup S'$ with $P_B(X|Y) - P_S(X|Y') \geq m - m' > \delta$. 
    
    

    Then a narrow bracketer facing simultaneous choices $S_1$ and $S_2$ will pick $X + Y - m$ from $S_1$ and $Y'$ from $S_2$ for a total outcome of $X + Y + Y' - m$. However, they could have chosen $Y$ from $S_1$ and $X + Y' - m'$ from $S_2$ for a total of $X + Y + Y' - m' = X + Y + Y' - m + (m - m')$, thus leaving an amount of $m - m' > \delta$ on the table, and the claim follows. 
\end{proof}

\paragraph{Proof of Proposition \ref{prop:price-representation}.}

\begin{proof}

Let $\bar{\Delta} \equiv \sup_{X, Y, Y'} P_B(X|Y) - P_S(X|Y')$ and suppose that $\bar{\Delta} = 0$. Then $P_B(X|Y) = P_S(X|Y')$ for all $X$ and $Y$, and hence $P_B(X|Y) = P_B(X|Y') \equiv P_B(X)$ for all $X$, $Y$, $Y'$. Moreover, $P_S(X|Y') = P_B(X)$ for all $Y'$, hence $P_S(X|Y') = P_S(X) = P_B(X)$. Hence we see that $P_B(X) = P_S(X) = P(X)$: the person always buys when the price of $X$ on top of the highest-value alternative is less than $P(X)$ and always sells when the price on top of the highest-value alternative is more than $P(X)$, and will do both when the price is $X$. Thus there is a unique price $P(X)$ for each $X$. 

Now first we show that if weak axiom of revealed preference (WARP) is violated, then it is violated by at least an amount $\varepsilon > 0$, meaning that there exist $X$, $Y$, $S$, and $S'$ s.t. $\{X, Y\} \subset S$, $\{X + \varepsilon, Y\} \subset S'$, with $X \in C(S)$ and $Y \in C(S')$, $X + \varepsilon \notin C(S')$, where $C$ is the choice correspondence. Note that for $\varepsilon = 0$, we have the usual definition of WARP.


Now suppose that we have a WARP violation. 
Let $X$, $Y$, $S$, and $S'$ be a WARP violation. We have that $X$, $Y \in S \cap S'$, with $X \in C(S)$ and $Y \in C(S')$ but $X \notin C(S')$. Note that if $X \in C(\{X, Y\})$, then there is also a WARP violation such that $S$ equals $\{X, Y\}$
Similarly, if $X \notin C(\{X, Y\})$, then we can set $S'$ equal $\{X, Y\}$.

\textbf{Case 1:} $X \in C(\{X, Y\})$ and $S = \{X, Y\}$.

First notice that if $C(\{X, Y\}) = \{X\}$, then there is some $\varepsilon > 0$ s.t. $C(\{X - \varepsilon, Y\}) = \{X - \varepsilon\}$ by continuity: if $Y$ was chosen for all sufficiently small $\varepsilon$, then it also would be chosen from $\{X, Y\}$. Thus in this case we have immediately a WARP violation of size $\varepsilon$. 

If $C(S') = \{Y\}$, then by continuity, we have that $C(S_{\varepsilon}^{'}) = \{Y\}$ where $S_{\varepsilon}^{'} = S \setminus \{X\} \cup \{X + \varepsilon\}$ for sufficiently small $\varepsilon$: if this was not the case, say $Z \neq Y, X + \varepsilon$ and $Z \in C(S_{\varepsilon}^{'})$, then by continuity $Z \in C(S')$ --- and a similar argument rules out that $X - \varepsilon$ is chosen. Thus in this case, we have now found a WARP violation of size $\varepsilon$.

If $C(S') \neq \{Y\}$, then consider $C(S_{\varepsilon}^{'})$: if this contains $Y$, then we immediately have a WARP violation of size $\varepsilon$. So suppose that $Y \notin C(S_{\varepsilon}^{'})$ for sufficiently small $\varepsilon$. Then there is some $Z \in C(S_{\varepsilon}^{'})$ that is not $X + \varepsilon$ (by continuity it can't be) nor $Y$. Now consider the set $\{Y, Z\}$. If $C(\{Y, Z\}) = \{Y\}$, then we can show as we did in the previous paragraph that there is some $\varepsilon > 0$ s.t. $C(\{Y - \varepsilon, Z\}) = \{Y - \varepsilon\}$. Therefore, since $Z \in C(S')$, we have found a WARP violation of size $\varepsilon$, since $Y - \varepsilon$ is chosen over $Z$ in one choice, yet $Z$ is chosen over $Y$ in the other. The same argument applies if $C(\{Y, Z\}) = \{Z\}$, since the situation is symmetric.

The remaining possibility is thus that $C(\{Y, Z\}) = \{Y, Z\}$. We also know that $Z$ is chosen from $S_{\varepsilon}^{'}$, while $Y$ and $X + \varepsilon$ are not. Thus this is an example of a WARP violation, since $Y$ is chosen from $\{Y, Z\}$, but only $Z$ is from $S_{\varepsilon}^{'}$. Moreover, our new example has choice $S$ of size $2$, and choice set $S_{\varepsilon}^{'}$ of same size as choice set $S'$. However, we can pick $\varepsilon$ s.t. the size of the choice set, $|C(S_{\varepsilon}^{'})|$, is at least one less than $|C(S')|$: by continuity, every option chosen from $S_{\varepsilon}^{'}$ will also be chosen in the limit as $\varepsilon \to 0$, moreover, there are a finite number of options that don't change with $\varepsilon$, other than $X - \varepsilon$, which we know is not chosen. Thus the size can at most be equal, but given that we know that $Y$ is not chosen for $\varepsilon > 0$, yet chosen for $\varepsilon = 0$, the size is at at least one less. 

But then we can repeat the previous argument for this newly found WARP violation and either we find a WARP violation of size $\varepsilon$ or a new violation where the size of $C(S')$ decreases again by at least one. Since this can never decrease to $0$, this process must eventually stop, which it only does once we find a WARP violation of size $\varepsilon$. 

\textbf{Case 2:} Suppose $C(\{X, Y\}) = \{Y\}$

In this case, the same argument as me made previously immediately applies that $C(\{X + \varepsilon, Y\}) = \{X + \varepsilon\}$ for sufficiently small $\varepsilon$ and hence we have a WARP violation of size $\varepsilon$.

Together these cases prove that if there is any WARP violation, we must have a WARP violation of size $\varepsilon$. Since a WARP violation of size $\varepsilon$ implies a price differential of size $\varepsilon$, this proves that $\Delta = 0$ implies that WARP is satisfied.

%
%
%
%

Hence we know that the choices can be represented by a transitive preference, $\succsim$. 
Consider the choice set where the person chooses between $Y$ and $Y + X - m$. Since $P(X|Y) = P(X)$, the person always chooses $Y$ over $X + Y - m$ for $m > P(X)$, and chooses only $X + Y - m$ over $Y$ for $m < P(X)$. Thus $X +Y - m \succ Y$ if and only if $m < P(X)$, $Y \succ X + Y - m$ if and only if $m > P(X)$, and thus by continuity $X + Y - P(X) \sim Y$. Letting $Y = Z + P(X)$, we see that $X + Z + P(X) - P(X) = X + Z \sim Z + P(X)$ for all $Z$ and $X$, hence setting $Z = 0$ we find that $X \sim P(X)$.

Since the person always chooses the strictly larger amount of money, when $P(X) > P(Y)$, $P(X)$ is strictly preferred to the amount $P(Y)$, so $P(X) \succ P(Y)$, and therefore $X \sim P(X) \succ P(Y) \sim Y$. Thus $X$ is strictly preferred to $Y$ if and only if $P(X) > P(Y)$. Thus $P(\cdot)$ is in fact a utility representation for the preference relation.

Moreover, since, as we showed above, $X + Z \sim P(X) + Z$, we also have that $P(X) + Z = Z + P(X) \sim P(Z) + P(X) = P(X) + P(Z)$. Further, since $X \sim P(X)$, we also have that $X + Z \sim P(X + Z)$, hence $P(X + Z) \sim X + Z \sim P(X) + Z \sim P(X) + P(Z)$. Finally, indifference between these two amounts of money implies that they are equal, i.e., if $P(X + Z) = P(X) + P(Z)$.

This completes the proof.
\end{proof}

\subsection{Implications of additive utility representations}\label{appendix:additive-utility}

We showed in the main text that $P(\cdot)$ is additive when $\Delta = 0$. The next Proposition formalizes what additivity implies across different choice domains. 

\begin{proposition}\label{prop:additivity-conditions}
  Suppose $P(\cdot): \mathbb{X} \to \mathbb{R}$ be an additive and continuous function. Then the following hold:
  \begin{enumerate}
    \item Let $\mathbb{X}$ be a space of random variables rich enough so that if $X$, $Y \in \mathbb{X}$, then there is some random variable $A$ distributed uniformly on $[0, 1]$ that is independent of $X$ and $Y$ with $\mathbbm{1}(A \in (p,q)) \cdot X + \mathbbm{1}(A \notin (p,q)) \cdot Y \in \mathbb{X}$ for any $p, q \in [0, 1]$. Assuming that $P(X) = P(Y)$ whenever $X$ and $Y$ have the same distribution, $P(\cdot)$ is the certainty equivalent for an expected-utility agent
    \item if $\mathbb{X}$ is a space of bounded real random variables, then $P(\cdot)$ is the certainty equivalent for a CARA agent
    \item if $\mathbb{X}$ is a space of bounded and independent real random variables, then $P(\cdot)$ is a weighted average of certainty equivalents of CARA agents
    \item if $\mathbb{X} = \mathbb{R}_{\geq 0}^{n}$ or $\mathbb{X} = \mathbb{R}^{n}$, then $P(\vect{x}) = \vect{x} \cdot \vect{\lambda}$ for some $\vect{\lambda} \in \mathbb{R}^{n}$ and for all $\vect{x} \in \mathbb{X}$
  \end{enumerate}
\end{proposition}

Except for the first result on expected utility, results 2 through 4 have been noted in separate papers as conditions under which bracketing unidentified: \cite{rabin2009narrow} assume expected utility and show that narrow bracketing incurs no cost (which is another way of stating that it is unidentified) if and only if agents have CARA preferences; \cite{ellisFreeman2020revealingBracketing} show that narrow bracketing is unidentified in multi-good choices when the utility is linear; and \cite{mu2021monotone} consider additive certainty equivalents like $P(\cdot)$ and prove statement 3 above. Our contribution is to show how additivity of $P(\cdot)$ is the unifying feature behind all these results, to highlight that it implies other conditions (such as on expected utility, result 1) and that it will generalize to all other settings.
We will now prove the special cases of Proposition \ref{prop:additivity-conditions}.

\begin{proof}
  \textbf{Case 1:} $\mathbb{X}$ is a (sufficiently rich) space of random variables

  Denote by $X$, $Y$, and $Z$ three random variables. Let $A$ be distributed uniformly on $[0,1]$, independently of $X$, $Y$, $Z$ and $A(p,q)$ be the event that $A \in (p, q)$. Then $\tilde{X} = \mathbbm{1}(A(0,p)) \cdot X + \mathbbm{1}(A(p,1)) \cdot Z$ and $\tilde{Y} = \mathbbm{1}(A(0,p)) \cdot Y + \mathbbm{1}(A(p,1) Z)$ are the random variables yielding the value of $X$ respectively $Y$ with probability $p$ and the value of $Z$ with probability $1 - p$.
  \begin{align*}
    P(\mathbbm{1}(A(0, p+q) \cdot X)) & = P(\mathbbm{1}(A(0,p)) \cdot X + \mathbbm{1}(A(p, p+q) \cdot X) \\
                                & = P(\mathbbm{1}(A(0,p)) \cdot X) + P(\mathbbm{1}(A(p, p+q) \cdot X) \text{, by additivity}\\
                                & =  P(\mathbbm{1}(A(0,p)) \cdot X) + P(\mathbbm{1}(A(0, q) \cdot X) \text{, by equal distributions }
  \end{align*}
  where we used the fact that $\mathbbm{1}(A(0,q)) \cdot X$ and $\mathbbm{1}(A(p,p+q)) \cdot X$ have the same distribution, hence also the same utility. Writing $f_{X}(p) = P(\mathbbm{1}(A(0,p)) \cdot X)$, we have that $f_{X}$ is additive, i.e. it satisfies $f_{X}(p + q) = f_{X}(p) + f_{X}(q)$. We assumed that it is continuous in $p$, hence we know that $f_{X}$ is linear, i.e. $f_{X}(p) = \lambda p$ for some $\lambda$. Since $f_{X}(1) = P(X)$, we have $\lambda = P(X)$, which shows that $P(\mathbbm{1}(A(0,p)) X) = p P(X)$.

  Let $\tilde{X} = \mathbbm{1}(A(0,p)) \cdot X + \mathbbm{1}(A(p,1)) \cdot Z$ and similarly for $\tilde{Y} = \mathbbm{1}(A(0,p)) \cdot Y + \mathbbm{1}(A(p,1)) \cdot Z$. Suppose that $X \sim Y$, so that $P(X) \geq P(Y)$. Then we have $P(\tilde{X}) = P(\mathbbm{1}(A(0,p)) \cdot X) + P(\mathbbm{1}(A(p,1)) \cdot Z) = p P(X) + (1 - p) P(Z) \geq p P(Y) + (1 - p) P(Z) = P(\mathbbm{1}(A(0,p)) \cdot Y) + P(\mathbbm{1}(A(p,1)) \cdot Z) = P(\tilde{Y})$. Hence $p \cdot X + (1 - p) \cdot Z \sim p \cdot Y + (1 - p) \cdot Z$ (the same argument applies to any event $B$ with probability $p$)  showing that the independence axiom holds for $\sim$. Together with continuity, this implies expected utiltiy.

  \textbf{Case 2:} $\mathbb{X}$ is a space of bounded real random variables

  From case 1 we know that we have expected utility preferences.

  Now let $X$ be any random variable and $w$ and $w' \in \mathbb{R}$. Then $X + w$ denotes the random variable yielding $w$ more than $X$. Then $P(X + w) = P(X) + P(w) \implies P(X + w) - P(w) = P(X) = P(X + w') - P(w')$. This is the certainty equivalent of $X$, once on top of $w$, once on top of $w'$, which has to be constant for all $w$ and $w'$. Hence we must have constant absolute risk aversion.

  \textbf{Case 3:} This result follows directly from Theorem 1 in \cite{mu2021monotone}, since $P(\cdot)$ in this context is a monotone additive statistic over bounded real-valued random variables.

  \textbf{Case 4:} $\mathbb{X} = \mathbb{R}^{n}$

  Letting $\vect{e}_{i}$ be the unit vector in dimension $i$, i.e. it is the bundle that provides one unit of good $i$ and nothing else, then we can define $f_{i}(x) := P(x \cdot e_{i})$ for $x \in \mathbb{R}$. $f_{i}(x + y) = P( (x + y) \cdot e_{i} ) = P( x \cdot e_{i} + y \cdot e_{i} ) = P( x \cdot e_{i}) + P( y \cdot e_{i} ) = f_{i}(x) + f_{i}(y)$, so $f_{i}$ is additive and continuous. It is well-known that additivity plus continuity for a function $f: \mathbb{R} \to \mathbb{R}$ implies that $f(\cdot)$ is linear.\footnote{Additivity defining a function is also called Cauchy's functional equation and the stated result dates back to Cauchy.} Thus $f_{i}(x) = \lambda_{i} \cdot x$ for some $\lambda_{i} \in \mathbb{R}$. By additivity, we have that $P(\vect{x}) = P( \sum_{i} x_{i} \cdot e_{i} ) = \sum_{i} P(x_{i} \cdot e_{i}) = \sum_{i} f_{i}(x_{i}) = \sum_{i} \lambda_{i} x_{i} = \vect{x} \cdot \vect{\lambda}$.
\end{proof}

%% file: appendix-instructions.tex
\section{Appendix: Instructions} \label{appendix:instructions}

\subsection{Welcome Screen}
Welcome \\
Thank you for accepting our HIT. \\
During the HIT, please do not close this window or leave the HIT's web pages in any other way. \\

If you do close your browser or leave the HIT, you will not be able to re-enter and we will not be able to pay you!  \\

You will receive a baseline payment of \$2.00 once you complete the HIT. Additionally, you can earn an extra bonus that will depend on your choices.\\

You will receive a code to enter into MTurk to collect your payment once you have finished.\\

Please read all instructions carefully.

\subsection{Instructions}

Thank you for accepting to participate in this HIT. On top of the guaranteed payment of \$2.00 you will have the chance to earn an extra bonus, as explained later.

The task
In this HIT you will decode several sequences of random letters into numbers with the given decoding table. For each letters sequence, the decoding table changes. The main part of the HIT will require you to decode several of these tasks.

To gain familiarity with the task you will now have to correctly decode 3 sequences. Note that each letter must be decoded correctly. After entering the decoded sequence, hit the submit button. Subsequently, irrespective of whether the text sequence was decoded correctly or not, a new sequence and decoding table will appear. Once you decode 3 sequences correctly, we describe the main part of the HIT.\\

\begin{figure}[H]
    \centering
\includegraphics[scale=0.8]{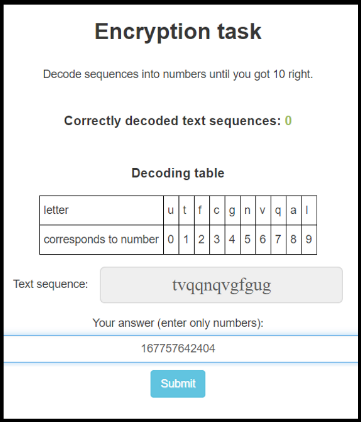}
    \caption{SCREENSHOT}
    \label{fig:task-instructions}
\end{figure}

In the example you see the text sequence \textit{tvqqnqvgfgug}. The decoding table tells you that u=0, t=1,... This means that you have to decode \textit{tvqqnqvgfgug} into \textit{167757642404} and enter this numeric value into the answer field.\\

\subsection{Task}

\subsection{Main Task instructions}

THE TASK

\textbf{[BOTH | MONEY/LOW | MONEY]} By completing the HIT you will receive \$2.00. To do so you are required to decode some sequences correctly for a bonus.\\

We will give you two pages of choices, with 16 choices on each page. Each choice is between a low number and a high number of \textbf{[additional - only in BOTH]} sequences to decode before the required sequences for different bonuses.\\

Example Choice (DOES NOT COUNT):\\

\begin{itemize}
    \item 10 \textbf{[additional - only in BOTH]} sequences for an extra \$4.00
    \item 20 \textbf{[additional - only in BOTH]} sequences for an extra \$5.00
\end{itemize}

After you made your choice the computer will select randomly one of the 16 choices from one of the 2 pages. That option will be implemented. Thus you should select your preferred option for each choice.

\textbf{[NONE]}
To complete the HIT you will be asked to decode a certain number of sequences correctly. The number of sequences you will be required to decode will depend on your choices.

We will give you two pages of choices, with 16 choices on each page. Each choice is between a low number and a high number of sequences to decode for different amounts (which includes the \$2.00 completion fee of the HIT).

\begin{itemize}
    \item 10 sequences for a total payment of \$6.00
    \item 20 sequences for a total payment of \$7.00
\end{itemize}

After you made your choice the computer will select randomly one of the 16 choices from one of the 2 pages. That option will be implemented. Thus you should select your preferred option for each choice.

\pagebreak

\subsection{Main Task - Scenario 1}\label{maintask}
\textbf{[NONE]}
\textbf{Note: you are required to decode 15 sequences correctly, in addition to the sequences based on your choices.}\\

\textbf{Choices to make now:} for each choice in this Scenario, choose the preferred option.\\
\textbf{By completing the HIT you will receive a total payment (which includes the \$2.00 completion fee) depending on your choices.}\\

\begin{center}\small{
\flushleft{\begin{tabular}{l l c c l}\hline
&& OPTION A & OPTION B & \\ \hline
1) & {\raggedright 15 sequences for a total payment of \$6.00} & $\bigcirc$ & $\bigcirc$ & {\raggedright 30 sequences for a total payment of \$6.25} \\
2) & {\raggedright 15 sequences for a total payment of \$6.00} & $\bigcirc$ & $\bigcirc$ & {\raggedright 30 sequences for a total payment of \$6.50} \\
3) & {\raggedright 15 sequences for a total payment of \$6.00} & $\bigcirc$ & $\bigcirc$ & {\raggedright 30 sequences for a total payment of \$6.75} \\
4) & {\raggedright 15 sequences for a total payment of \$6.00} & $\bigcirc$ & $\bigcirc$ & {\raggedright 30 sequences for a total payment of \$7.00} \\
5) & {\raggedright 15 sequences for a total payment of \$6.00} & $\bigcirc$ & $\bigcirc$ & {\raggedright 30 sequences for a total payment of \$7.25} \\
6) & {\raggedright 15 sequences for a total payment of \$6.00} & $\bigcirc$ & $\bigcirc$ & {\raggedright 30 sequences for a total payment of \$7.50} \\
7) & {\raggedright 15 sequences for a total payment of \$6.00} & $\bigcirc$ & $\bigcirc$ & {\raggedright 30 sequences for a total payment of \$7.75} \\
8) & {\raggedright 15 sequences for a total payment of \$6.00} & $\bigcirc$ & $\bigcirc$ & {\raggedright 30 sequences for a total payment of \$8.00} \\
9) & {\raggedright 15 sequences for a total payment of \$6.00} & $\bigcirc$ & $\bigcirc$ & {\raggedright 30 sequences for a total payment of \$8.25} \\
10) & {\raggedright 15 sequences for a total payment of \$6.00} & $\bigcirc$ & $\bigcirc$ & {\raggedright 30 sequences for a total payment of \$8.50} \\
11) & {\raggedright 15 sequences for a total payment of \$6.00} & $\bigcirc$ & $\bigcirc$ & {\raggedright 30 sequences for a total payment of \$8.75} \\
12) & {\raggedright 15 sequences for a total payment of \$6.00} & $\bigcirc$ & $\bigcirc$ & {\raggedright 30 sequences for a total payment of \$9.00} \\
13) & {\raggedright 15 sequences for a total payment of \$6.00} & $\bigcirc$ & $\bigcirc$ & {\raggedright 30 sequences for a total payment of \$9.25} \\
14) & {\raggedright 15 sequences for a total payment of \$6.00} & $\bigcirc$ & $\bigcirc$ & {\raggedright 30 sequences for a total payment of \$9.50} \\
15) & {\raggedright 15 sequences for a total payment of \$6.00} & $\bigcirc$ & $\bigcirc$ & {\raggedright 30 sequences for a total payment of \$9.75} \\
16) & {\raggedright 15 sequences for a total payment of \$6.00} & $\bigcirc$ & $\bigcirc$ & {\raggedright 30 sequences for a total payment of \$10.00} \\
\hline
\end{tabular}}}
\end{center}

\pagebreak

\textbf{[MONEY]}
\textbf{Note: you are required to decode 15 sequences correctly, in addition to the sequences based on your choices.}\\

\textbf{Choices to make now:} for each choice in this Scenario, choose the preferred option.\\
\textbf{By completing the HIT you will receive \$2.00 plus a bonus depending on your choices.}\\

\begin{center}\small{
\flushleft{\begin{tabular}{l l c c l}\hline
&& OPTION A & OPTION B & \\ \hline
1) & {\raggedright 15 sequences for an extra of \$4.00} & $\bigcirc$ & $\bigcirc$ & {\raggedright 30 sequences for an extra of \$4.25} \\
2) & {\raggedright 15 sequences for an extra of \$4.00} & $\bigcirc$ & $\bigcirc$ & {\raggedright 30 sequences for an extra of \$4.50} \\
3) & {\raggedright 15 sequences for an extra of \$4.00} & $\bigcirc$ & $\bigcirc$ & {\raggedright 30 sequences for an extra of \$4.75} \\
4) & {\raggedright 15 sequences for an extra of \$4.00} & $\bigcirc$ & $\bigcirc$ & {\raggedright 30 sequences for an extra of \$5.00} \\
5) & {\raggedright 15 sequences for an extra of \$4.00} & $\bigcirc$ & $\bigcirc$ & {\raggedright 30 sequences for an extra of \$5.25} \\
6) & {\raggedright 15 sequences for an extra of \$4.00} & $\bigcirc$ & $\bigcirc$ & {\raggedright 30 sequences for an extra of \$5.50} \\
7) & {\raggedright 15 sequences for an extra of \$4.00} & $\bigcirc$ & $\bigcirc$ & {\raggedright 30 sequences for an extra of \$5.75} \\
8) & {\raggedright 15 sequences for an extra of \$4.00} & $\bigcirc$ & $\bigcirc$ & {\raggedright 30 sequences for an extra of \$6.00} \\
9) & {\raggedright 15 sequences for an extra of \$4.00} & $\bigcirc$ & $\bigcirc$ & {\raggedright 30 sequences for an extra of \$6.25} \\
10) & {\raggedright 15 sequences for an extra of \$4.00} & $\bigcirc$ & $\bigcirc$ & {\raggedright 30 sequences for an extra of \$6.50} \\
11) & {\raggedright 15 sequences for an extra of \$4.00} & $\bigcirc$ & $\bigcirc$ & {\raggedright 30 sequences for an extra of \$6.75} \\
12) & {\raggedright 15 sequences for an extra of \$4.00} & $\bigcirc$ & $\bigcirc$ & {\raggedright 30 sequences for an extra of \$7.00} \\
13) & {\raggedright 15 sequences for an extra of \$4.00} & $\bigcirc$ & $\bigcirc$ & {\raggedright 30 sequences for an extra of \$7.25} \\
14) & {\raggedright 15 sequences for an extra of \$4.00} & $\bigcirc$ & $\bigcirc$ & {\raggedright 30 sequences for an extra of \$7.50} \\
15) & {\raggedright 15 sequences for an extra of \$4.00} & $\bigcirc$ & $\bigcirc$ & {\raggedright 30 sequences for an extra of \$7.75} \\
16) & {\raggedright 15 sequences for an extra of \$4.00} & $\bigcirc$ & $\bigcirc$ & {\raggedright 30 sequences for an extra of \$8.00} \\
\hline
\end{tabular}}}
\end{center}

\pagebreak

\textbf{[BOTH]}

\textbf{Note: you are required to decode 15 sequences correctly, in addition to the sequences based on your choices.}\\
\textbf{Choices to make now:} for each choice in this Scenario, choose the preferred option.\\
\textbf{By completing the HIT you will receive \$2.00 plus a bonus depending on your choices.}\\

\begin{center}\small{
\flushleft{\begin{tabular}{l l c c l}\hline
&& OPTION A & OPTION B & \\ \hline
1) & {\raggedright 0 additional sequences for an extra \$4.00} & $\bigcirc$ & $\bigcirc$ & {\raggedright 15 additional sequences for an extra \$4.25} \\
2) & {\raggedright 0 additional sequences for an extra \$4.00} & $\bigcirc$ & $\bigcirc$ & {\raggedright 15 additional sequences for an extra \$4.50} \\
3) & {\raggedright 0 additional sequences for an extra \$4.00} & $\bigcirc$ & $\bigcirc$ & {\raggedright 15 additional sequences for an extra \$4.75} \\
4) & {\raggedright 0 additional sequences for an extra \$4.00} & $\bigcirc$ & $\bigcirc$ & {\raggedright 15 additional sequences for an extra \$5.00} \\
5) & {\raggedright 0 additional sequences for an extra \$4.00} & $\bigcirc$ & $\bigcirc$ & {\raggedright 15 additional sequences for an extra \$5.25} \\
6) & {\raggedright 0 additional sequences for an extra \$4.00} & $\bigcirc$ & $\bigcirc$ & {\raggedright 15 additional sequences for an extra \$5.50} \\
7) & {\raggedright 0 additional sequences for an extra \$4.00} & $\bigcirc$ & $\bigcirc$ & {\raggedright 15 additional sequences for an extra \$5.75} \\
8) & {\raggedright 0 additional sequences for an extra \$4.00} & $\bigcirc$ & $\bigcirc$ & {\raggedright 15 additional sequences for an extra \$6.00} \\
9) & {\raggedright 0 additional sequences for an extra \$4.00} & $\bigcirc$ & $\bigcirc$ & {\raggedright 15 additional sequences for an extra \$6.25} \\
10) & {\raggedright 0 additional sequences for an extra \$4.00} & $\bigcirc$ & $\bigcirc$ & {\raggedright 15 additional sequences for an extra \$6.50} \\
11) & {\raggedright 0 additional sequences for an extra \$4.00} & $\bigcirc$ & $\bigcirc$ & {\raggedright 15 additional sequences for an extra \$6.75} \\
12) & {\raggedright 0 additional sequences for an extra \$4.00} & $\bigcirc$ & $\bigcirc$ & {\raggedright 15 additional sequences for an extra \$7.00} \\
13) & {\raggedright 0 additional sequences for an extra \$4.00} & $\bigcirc$ & $\bigcirc$ & {\raggedright 15 additional sequences for an extra \$7.25} \\
14) & {\raggedright 0 additional sequences for an extra \$4.00} & $\bigcirc$ & $\bigcirc$ & {\raggedright 15 additional sequences for an extra \$7.50} \\
15) & {\raggedright 0 additional sequences for an extra \$4.00} & $\bigcirc$ & $\bigcirc$ & {\raggedright 15 additional sequences for an extra \$7.75} \\
16) & {\raggedright 0 additional sequences for an extra \$4.00} & $\bigcirc$ & $\bigcirc$ & {\raggedright 15 additional sequences for an extra \$8.00} \\
\hline
\end{tabular}}}
\end{center}

\pagebreak

\textbf{[MONEY/LOW]}

\textbf{Choices to make now:} for each choice in this Scenario, choose the preferred option.\\
\textbf{By completing the HIT you will receive \$2.00 plus a bonus depending on your choices.}\\

\begin{center}\small{
\flushleft{\begin{tabular}{l l c c l}\hline
&& OPTION A & OPTION B & \\ \hline
1) & {\raggedright 0 sequences for an extra of \$4.00} & $\bigcirc$ & $\bigcirc$ & {\raggedright 15 sequences for an extra of \$4.25} \\
2) & {\raggedright 0 sequences for an extra of \$4.00} & $\bigcirc$ & $\bigcirc$ & {\raggedright 15 sequences for an extra of \$4.50} \\
3) & {\raggedright 0 sequences for an extra of \$4.00} & $\bigcirc$ & $\bigcirc$ & {\raggedright 15 sequences for an extra of \$4.75} \\
4) & {\raggedright 0 sequences for an extra of \$4.00} & $\bigcirc$ & $\bigcirc$ & {\raggedright 15 sequences for an extra of \$5.00} \\
5) & {\raggedright 0 sequences for an extra of \$4.00} & $\bigcirc$ & $\bigcirc$ & {\raggedright 15 sequences for an extra of \$5.25} \\
6) & {\raggedright 0 sequences for an extra of \$4.00} & $\bigcirc$ & $\bigcirc$ & {\raggedright 15 sequences for an extra of \$5.50} \\
7) & {\raggedright 0 sequences for an extra of \$4.00} & $\bigcirc$ & $\bigcirc$ & {\raggedright 15 sequences for an extra of \$5.75} \\
8) & {\raggedright 0 sequences for an extra of \$4.00} & $\bigcirc$ & $\bigcirc$ & {\raggedright 15 sequences for an extra of \$6.00} \\
9) & {\raggedright 0 sequences for an extra of \$4.00} & $\bigcirc$ & $\bigcirc$ & {\raggedright 15 sequences for an extra of \$6.25} \\
10) & {\raggedright 0 sequences for an extra of \$4.00} & $\bigcirc$ & $\bigcirc$ & {\raggedright 15 sequences for an extra of \$6.50} \\
11) & {\raggedright 0 sequences for an extra of \$4.00} & $\bigcirc$ & $\bigcirc$ & {\raggedright 15 sequences for an extra of \$6.75} \\
12) & {\raggedright 0 sequences for an extra of \$4.00} & $\bigcirc$ & $\bigcirc$ & {\raggedright 15 sequences for an extra of \$7.00} \\
13) & {\raggedright 0 sequences for an extra of \$4.00} & $\bigcirc$ & $\bigcirc$ & {\raggedright 15 sequences for an extra of \$7.25} \\
14) & {\raggedright 0 sequences for an extra of \$4.00} & $\bigcirc$ & $\bigcirc$ & {\raggedright 15 sequences for an extra of \$7.50} \\
15) & {\raggedright 0 sequences for an extra of \$4.00} & $\bigcirc$ & $\bigcirc$ & {\raggedright 15 sequences for an extra of \$7.75} \\
16) & {\raggedright 0 sequences for an extra of \$4.00} & $\bigcirc$ & $\bigcirc$ & {\raggedright 15 sequences for an extra of \$8.00} \\
\hline
\end{tabular}}}
\end{center}

\pagebreak

\subsection{Main Task - Scenario 2}

\textbf{[NONE]}

\textbf{Choices to make now:} for each choice in this Scenario, choose the preferred option.\\
\textbf{By completing the HIT you will receive a total payment (which includes the \$2.00 completion fee) depending on your choices.}\\

\begin{center}\small{
\flushleft{\begin{tabular}{l l c c l}\hline
&& OPTION A & OPTION B & \\ \hline
1) & {\raggedright 30 sequences for a total payment of \$6.00} & $\bigcirc$ & $\bigcirc$ & {\raggedright 45 sequences for a total payment of \$6.25} \\
2) & {\raggedright 30 sequences for a total payment of \$6.00} & $\bigcirc$ & $\bigcirc$ & {\raggedright 45 sequences for a total payment of \$6.50} \\
3) & {\raggedright 30 sequences for a total payment of \$6.00} & $\bigcirc$ & $\bigcirc$ & {\raggedright 45 sequences for a total payment of \$6.75} \\
4) & {\raggedright 30 sequences for a total payment of \$6.00} & $\bigcirc$ & $\bigcirc$ & {\raggedright 45 sequences for a total payment of \$7.00} \\
5) & {\raggedright 30 sequences for a total payment of \$6.00} & $\bigcirc$ & $\bigcirc$ & {\raggedright 45 sequences for a total payment of \$7.25} \\
6) & {\raggedright 30 sequences for a total payment of \$6.00} & $\bigcirc$ & $\bigcirc$ & {\raggedright 45 sequences for a total payment of \$7.50} \\
7) & {\raggedright 30 sequences for a total payment of \$6.00} & $\bigcirc$ & $\bigcirc$ & {\raggedright 45 sequences for a total payment of \$7.75} \\
8) & {\raggedright 30 sequences for a total payment of \$6.00} & $\bigcirc$ & $\bigcirc$ & {\raggedright 45 sequences for a total payment of \$8.00} \\
9) & {\raggedright 30 sequences for a total payment of \$6.00} & $\bigcirc$ & $\bigcirc$ & {\raggedright 45 sequences for a total payment of \$8.25} \\
10) & {\raggedright 30 sequences for a total payment of \$6.00} & $\bigcirc$ & $\bigcirc$ & {\raggedright 45 sequences for a total payment of \$8.50} \\
11) & {\raggedright 30 sequences for a total payment of \$6.00} & $\bigcirc$ & $\bigcirc$ & {\raggedright 45 sequences for a total payment of \$8.75} \\
12) & {\raggedright 30 sequences for a total payment of \$6.00} & $\bigcirc$ & $\bigcirc$ & {\raggedright 45 sequences for a total payment of \$9.00} \\
13) & {\raggedright 30 sequences for a total payment of \$6.00} & $\bigcirc$ & $\bigcirc$ & {\raggedright 45 sequences for a total payment of \$9.25} \\
14) & {\raggedright 30 sequences for a total payment of \$6.00} & $\bigcirc$ & $\bigcirc$ & {\raggedright 45 sequences for a total payment of \$9.50} \\
15) & {\raggedright 30 sequences for a total payment of \$6.00} & $\bigcirc$ & $\bigcirc$ & {\raggedright 45 sequences for a total payment of \$9.75} \\
16) & {\raggedright 30 sequences for a total payment of \$6.00} & $\bigcirc$ & $\bigcirc$ & {\raggedright 45 sequences for a total payment of \$10.00} \\
\hline
\end{tabular}}}
\end{center}

\pagebreak

\textbf{[BOTH]}

\textbf{Note: you are required to decode 15 sequences correctly, in addition to the sequences based on your choices.}\\
\textbf{Choices to make now:} for each choice in this Scenario, choose the preferred option.\\
\textbf{By completing the HIT you will receive \$2.00 plus a bonus depending on your choices.}\\

\begin{center}\small{
\flushleft{\begin{tabular}{l l c c l}\hline
&& OPTION A & OPTION B & \\ \hline
1) & {\raggedright 15 additional sequences for an extra \$4.00} & $\bigcirc$ & $\bigcirc$ & {\raggedright 30 additional sequences for an extra \$4.25} \\
2) & {\raggedright 15 additional sequences for an extra \$4.00} & $\bigcirc$ & $\bigcirc$ & {\raggedright 30 additional sequences for an extra \$4.50} \\
3) & {\raggedright 15 additional sequences for an extra \$4.00} & $\bigcirc$ & $\bigcirc$ & {\raggedright 30 additional sequences for an extra \$4.75} \\
4) & {\raggedright 15 additional sequences for an extra \$4.00} & $\bigcirc$ & $\bigcirc$ & {\raggedright 30 additional sequences for an extra \$5.00} \\
5) & {\raggedright 15 additional sequences for an extra \$4.00} & $\bigcirc$ & $\bigcirc$ & {\raggedright 30 additional sequences for an extra \$5.25} \\
6) & {\raggedright 15 additional sequences for an extra \$4.00} & $\bigcirc$ & $\bigcirc$ & {\raggedright 30 additional sequences for an extra \$5.50} \\
7) & {\raggedright 15 additional sequences for an extra \$4.00} & $\bigcirc$ & $\bigcirc$ & {\raggedright 30 additional sequences for an extra \$5.75} \\
8) & {\raggedright 15 additional sequences for an extra \$4.00} & $\bigcirc$ & $\bigcirc$ & {\raggedright 30 additional sequences for an extra \$6.00} \\
9) & {\raggedright 15 additional sequences for an extra \$4.00} & $\bigcirc$ & $\bigcirc$ & {\raggedright 30 additional sequences for an extra \$6.25} \\
10) & {\raggedright 15 additional sequences for an extra \$4.00} & $\bigcirc$ & $\bigcirc$ & {\raggedright 30 additional sequences for an extra \$6.50} \\
11) & {\raggedright 15 additional sequences for an extra \$4.00} & $\bigcirc$ & $\bigcirc$ & {\raggedright 30 additional sequences for an extra \$6.75} \\
12) & {\raggedright 15 additional sequences for an extra \$4.00} & $\bigcirc$ & $\bigcirc$ & {\raggedright 30 additional sequences for an extra \$7.00} \\
13) & {\raggedright 15 additional sequences for an extra \$4.00} & $\bigcirc$ & $\bigcirc$ & {\raggedright 30 additional sequences for an extra \$7.25} \\
14) & {\raggedright 15 additional sequences for an extra \$4.00} & $\bigcirc$ & $\bigcirc$ & {\raggedright 30 additional sequences for an extra \$7.50} \\
15) & {\raggedright 15 additional sequences for an extra \$4.00} & $\bigcirc$ & $\bigcirc$ & {\raggedright 30 additional sequences for an extra \$7.75} \\
16) & {\raggedright 15 additional sequences for an extra \$4.00} & $\bigcirc$ & $\bigcirc$ & {\raggedright 30 additional sequences for an extra \$8.00} \\
\hline
\end{tabular}}}
\end{center}

\pagebreak

\textbf{[MONEY/LOW]}

\textbf{Choices to make now:} for each choice in this Scenario, choose the preferred option.\\
\textbf{By completing the HIT you will receive \$2.00 plus a bonus depending on your choices.}\\

\begin{center}\small{
\flushleft{\begin{tabular}{l l c c l}\hline
&& OPTION A & OPTION B & \\ \hline
1) & {\raggedright 15 sequences for an extra bonus of \$4.00} & $\bigcirc$ & $\bigcirc$ & {\raggedright 30 sequences for an extra bonus of \$4.25} \\
2) & {\raggedright 15 sequences for an extra bonus of \$4.00} & $\bigcirc$ & $\bigcirc$ & {\raggedright 30 sequences for an extra bonus of \$4.50} \\
3) & {\raggedright 15 sequences for an extra bonus of \$4.00} & $\bigcirc$ & $\bigcirc$ & {\raggedright 30 sequences for an extra bonus of \$4.75} \\
4) & {\raggedright 15 sequences for an extra bonus of \$4.00} & $\bigcirc$ & $\bigcirc$ & {\raggedright 30 sequences for an extra bonus of \$5.00} \\
5) & {\raggedright 15 sequences for an extra bonus of \$4.00} & $\bigcirc$ & $\bigcirc$ & {\raggedright 30 sequences for an extra bonus of \$5.25} \\
6) & {\raggedright 15 sequences for an extra bonus of \$4.00} & $\bigcirc$ & $\bigcirc$ & {\raggedright 30 sequences for an extra bonus of \$5.50} \\
7) & {\raggedright 15 sequences for an extra bonus of \$4.00} & $\bigcirc$ & $\bigcirc$ & {\raggedright 30 sequences for an extra bonus of \$5.75} \\
8) & {\raggedright 15 sequences for an extra bonus of \$4.00} & $\bigcirc$ & $\bigcirc$ & {\raggedright 30 sequences for an extra bonus of \$6.00} \\
9) & {\raggedright 15 sequences for an extra bonus of \$4.00} & $\bigcirc$ & $\bigcirc$ & {\raggedright 30 sequences for an extra bonus of \$6.25} \\
115) & {\raggedright 15 sequences for an extra bonus of \$4.00} & $\bigcirc$ & $\bigcirc$ & {\raggedright 30 sequences for an extra bonus of \$6.50} \\
11) & {\raggedright 15 sequences for an extra bonus of \$4.00} & $\bigcirc$ & $\bigcirc$ & {\raggedright 30 sequences for an extra bonus of \$6.75} \\
12) & {\raggedright 15 sequences for an extra bonus of \$4.00} & $\bigcirc$ & $\bigcirc$ & {\raggedright 30 sequences for an extra bonus of \$7.00} \\
13) & {\raggedright 15 sequences for an extra bonus of \$4.00} & $\bigcirc$ & $\bigcirc$ & {\raggedright 30 sequences for an extra bonus of \$7.25} \\
14) & {\raggedright 15 sequences for an extra bonus of \$4.00} & $\bigcirc$ & $\bigcirc$ & {\raggedright 30 sequences for an extra bonus of \$7.50} \\
15) & {\raggedright 15 sequences for an extra bonus of \$4.00} & $\bigcirc$ & $\bigcirc$ & {\raggedright 30 sequences for an extra bonus of \$7.75} \\
16) & {\raggedright 15 sequences for an extra bonus of \$4.00} & $\bigcirc$ & $\bigcirc$ & {\raggedright 30 sequences for an extra bonus of \$8.00} \\
\hline
\end{tabular}}}
\end{center}

\pagebreak

\subsection{Results}

\textbf{[BOTH]}
SUMMARY OF THE TASK 

The computer randomly selected the Choice \# from Scenario \#. \\

For this option you selected that you are (are not) willing to decode \# additional sequences for \$X.XX. \\

In total you will decode \# sequences to receive the HIT payment and the bonus.\\

\textbf{[NONE \& MONEY/LOW]}
SUMMARY OF THE TASK 

The computer randomly selected the Choice \# from Scenario \#. \\

For this option you selected that you are (are not) willing to decode \# sequences in total for \$X.XX. \\

In total you will decode \# sequences to receive the HIT payment and the bonus. \\

\subsection{MAIN TASK}
\subsection{PAYMENT PAGE}
Your earnings\\
In today's HIT you have earned a bonus of $\$$.\\ 
Your guaranteed participation fee is: $\$2.00$.\\ 
So, in total, you have earned $\$$.\\

To receive your earnings, please enter this code into MTurk\\
After you have done that, you can close this window. We thank you for participating in our study.

\pagebreak

%% file: appendix-additional-results.tex
\section{Appendix: Additional Results}\label{appendix:additional-results}

\subsection{Main Results: Wilcoxon tests, tests by gender}\label{appendix:wilcox-and-gender}

Here we report results comparing the treatments by Wilcoxon and t-tests, running them for all participants, as well as for female and male participants separately.

\import{./data/}{mwu_all_tables.tex}

\import{./data/}{t_test_all_tables.tex}

\pagebreak

\subsection{Implementation Details}\label{appendix:implementation-details}

We collected more participants in the BOTH treatment than in other treatments, because our initial version of the BOTH treatment informed participants of their endowment on the page \emph{before} the first choice. In later versions we informed participants of their endowment only on the first choice page to (fast) reference effects not present in other treatments, where participants saw the information only on the first choice page. For this reason we continued collecting observations in this version of BOTH until we had enough data to compare it to the other treatments, which is why we ended up with more observations in BOTH (pooling all versions) than for the other treatments. As we show in Appendix \ref{appendix:reveal-before-page}, our results are robust to this change.

Other minor details that changed between our pre-registration and our actual implementation are that, fixing total outcomes, we only used two levels for the endowment (0 and 15 tasks) instead of three (0, 8, and 16 tasks); and that we limited ourselves to 2 scenarios per person rather than 5.

We added the treatment MONEY later to disentangle the contribution of bracketing of money and work dimensions separately. Specifically, the MONEY treatment combines all the tasks, hence differences between NONE and MONEY cannot be driven by a failure to combine work, while differences between MONEY and BOTH are direct evidence of narrow bracketing of money. To distinguish fully between the two, we should also have included a treatment that combines money but not work. This version of the MONEY treatment has the benefit that it can directly be compared to MONEY/LOW, providing additional tests of narrow bracketing.

The following table \ref{tab:session_summary} shows how many participants we recruited in which treatments during sessions on different days.

\import{./data/}{session_summary.tex}

\pagebreak

\subsection{Attrition}

Table \ref{tab:attrition} displays at what stage participants dropped out of the study.

\import{./data/}{attrition_statistics.tex}

\pagebreak

\subsection{Main treatments balanced data only}\label{appendix:balanced}

Since our experimental sessions were not always balanced, one possible concern might be that we get different results due to different populations across sessions. To alleviate this concern with respect to the main treatments (NONE, BOTH, and MONEY/LOW), we report here the differences in t-tests and Wilcoxon tests when restricting ourselves to the data that was collected in a balanced session -- that is, data where membership was randomized and equal within each session. As the results show, we still reject broad bracketing, and fail to reject narrow bracketing.

\import{./data/}{t_test_balanced.tex}

\import{./data/}{mwu_balanced.tex}

\pagebreak

\subsection{Baseline Tasks revealed right before the choice page only}\label{appendix:reveal-before-page}

Next we report results from the initial BOTH treatments where the information was displayed on the page right before. The results are essentially the same, although there is no longer a statistically significant difference in scenario 2, since BOTH lies between MONEY/LOW and NONE and is not significantly different from either, reflecting the lower power due to closer to 'linear' preferences in Scenario 2 (the difference between MONEY/LOW and NONE is lower).

\import{./data/}{t_test_presentation_before_page.tex}

We compare the means of BOTH treatments with message displayed before the first choice page and on the first choice page by scenario directly in Table \ref{tab:t_test_narrow_v_narrow}. This shows that for scenario 2, these two versions are significantly (and sizeably) different, reflecting also that in one case this leads to rejection of broad bracketing in scenario 2 and once it doesn't. No matter which is the accurate treatment, both reject broad bracketing, and neither rejects narrow bracketing.

There are two possible reasons for the difference: either it is due to the display of information, in which case the later data with information display on the page is the appropriate test, rejecting broad in both scenarios. In this case, the treatments BOTH and NONE are not balanced within sessions, since we had completed collection of data on NONE (mostly at least, we have a small overlap between the treatments). Or it is due to changes in the population due to sampling at different times. In this case the earlier data is the appropriate test, and balances observations against the NONE treatment -- i.e. the rejection of broad bracketing cannot be due (or more correctly, is statistically unlikely to be due) to different preferences.

\import{./data/}{t_test_narrow_v_narrow.tex}

\pagebreak

\subsection{Individual-level changes in reservation wage between scenarios}\label{subsec:change_in_rw_across_scenarios}

Frequencies of individuals who switched up, down, or stayed, and size of jumps by switching up or down.

\import{./data/}{rw_change_between_scenarios.tex}

\import{./data/}{rw_direction_change_between_scenarios.tex}

\pagebreak

\subsection{Plots of reservation wages by treatment}\label{raw_data_plots}

\begin{figure}[H]
  \caption{A bar plot of the raw reservation wages by treatment and scenario}
  \includegraphics[width=0.95\textwidth]{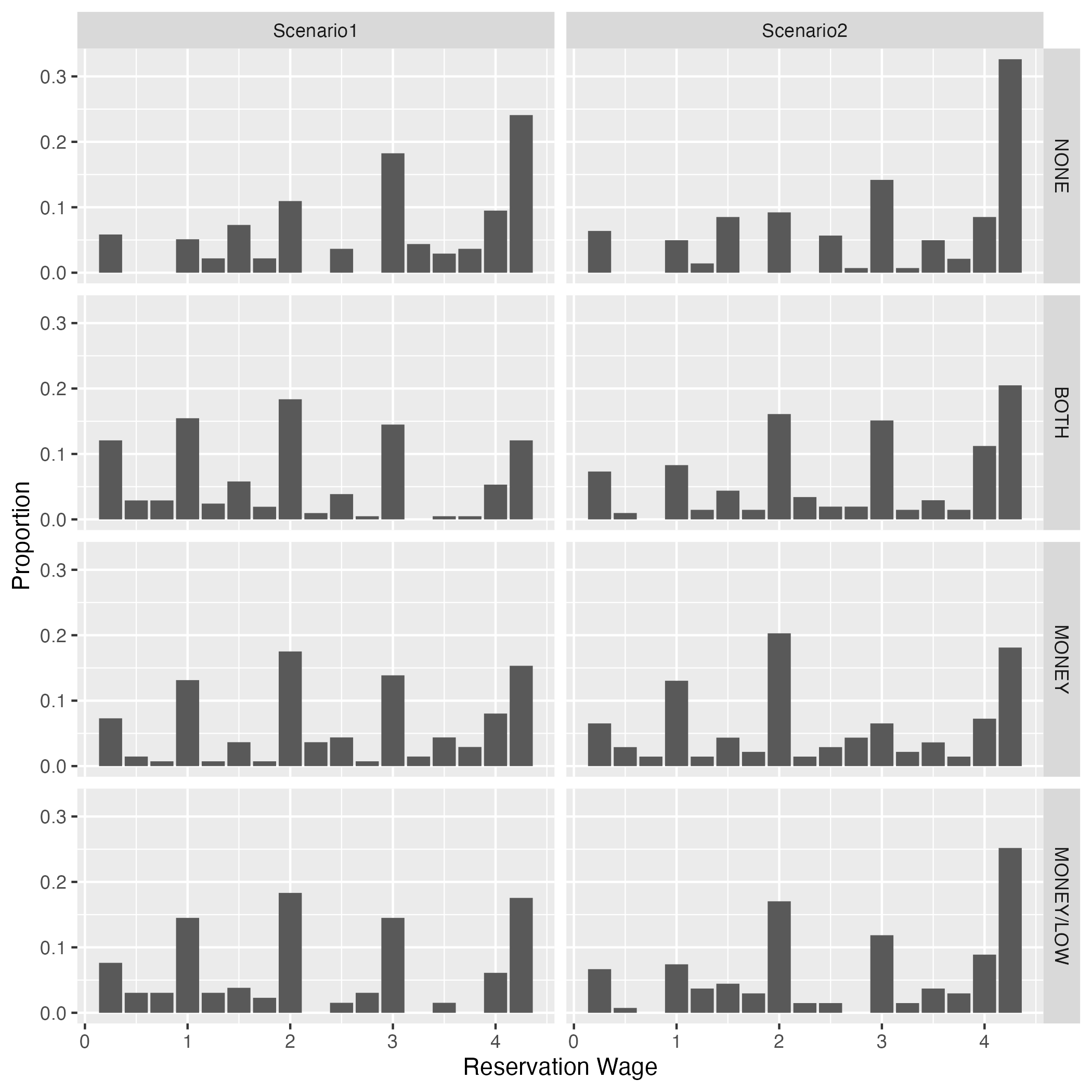}
\end{figure}

\begin{figure}[H]
  \caption{A kernel density plot of the raw reservation wages by treatment and scenario}
  \includegraphics[width=0.95\textwidth]{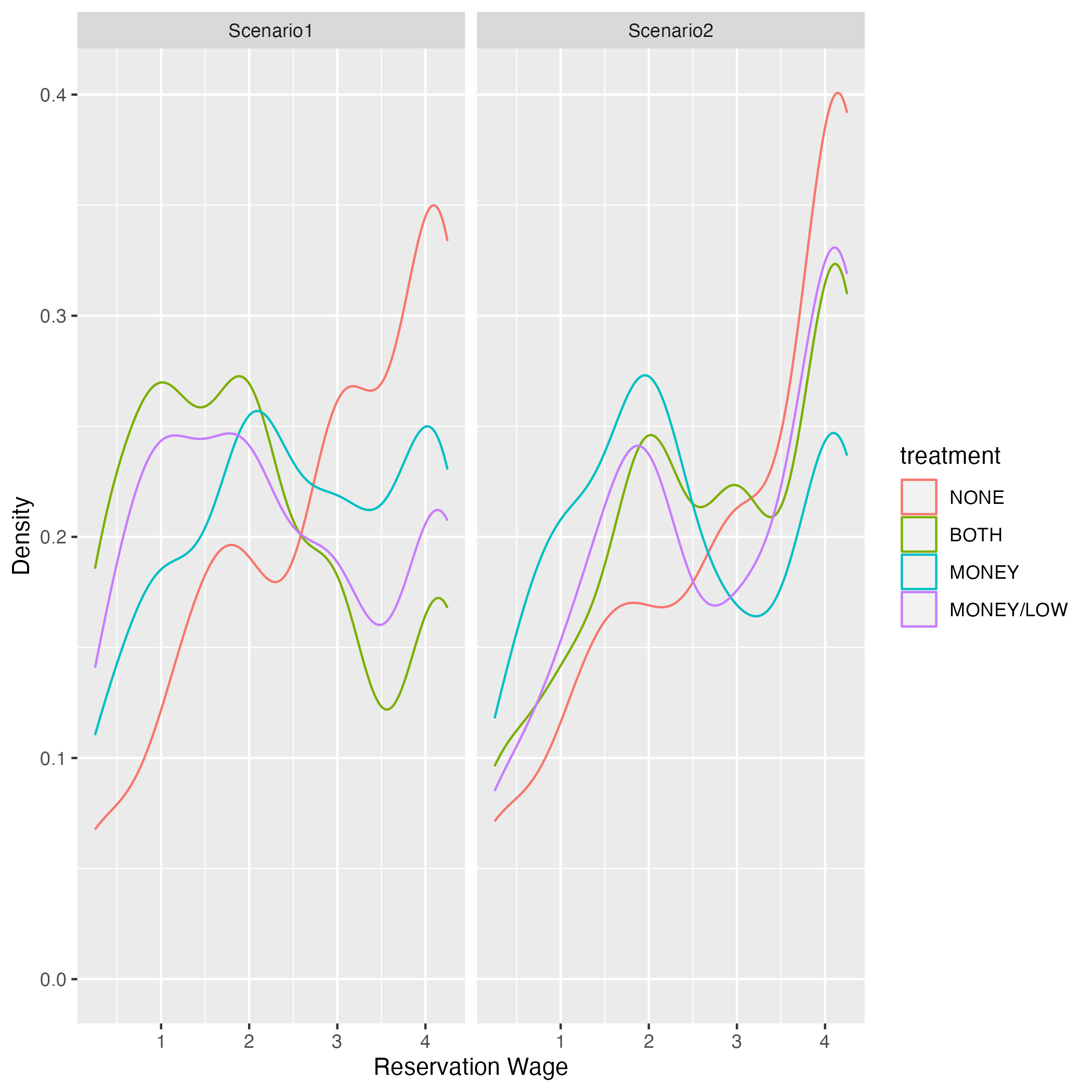}
\end{figure}

\pagebreak

\subsection{Linear Regression Results}\label{appendix:linear-regression}

In our pre-registration, we stated that we would use a linear regression that averages across the scenarios to test for bracketing, whether broad or narrow. This is however not the appropriate test, since bracketing should apply for each Scenario separately, not just for the average scenario, which is why those are the tests we use instead. This is best illustrated via an actual example we have in our data. We report the linear regression results in \ref{tab:linear-regressions}, which rejects broad bracketing overall; fails to reject narrow bracketing; and rejects broad bracketing of money alone. It does not however reject broad bracketing of work alone, since the treatment effects for MONEY and BOTH are $-0.47$ and $-0.55$ respectively with standard errors of around $0.11$, so they are not different.

\import{./data/}{linear_regression_pooled_and_by_scenario.tex}

But, as the results from \ref{tab:means_work_bracketing} shows, this is because the treatment effect in Scenario 1 is substantially smaller for BOTH than for MONEY ($2.07$ vs $2.50$); yet in Scenario 2, it is substantially larger for BOTH than for MONEY ($2.70$ vs $2.43$) which averages out to a statistically insignificant difference of $0.08$ when averaged across scenarios. Since narrow and broad bracketing make predictions for every choice decision, their predictions hold for each scenario individually, thus the linear regression is not the appropriate test and overly conservative.

\import{./data/}{means_work_bracketing.tex}

The fact that the reservation wage in MONEY stays the same across the two scenarios is consistent with participants bracketing all their endowments narrowly: the reservation wage for NONE is also almost constant, $2.88$ in Scenario 1 vs $2.99$ in Scenario 2 (see \ref{tab:means_data}). Our data suggests that doing 15 more tasks is equally unpleasant when one has to do 15 tasks or 30 tasks, both in treatments NONE and MONEY. The lower reservation wage in MONEY reflects the fact that MONEY has an endowment of money. If they ignore it and have concave utility from money, then for the same experienced disutility of work, they should be more willing to work for money in MONEY than in NONE. In the latter, they take into account that their wealth is higher.

\import{./data/}{means_data.tex}

\subsection{Size of reservation wage changes due to bracketing}

Going beyond our pre-registration, we estimate how much the reservation wage of participants changes due to bracketing. We report these changes and their standard errors in Table \ref{tab:cost_of_bracketing}. As we can see, the reservation wages change by as much as $\$0.82$. We also provide the equivalent of this change in reservation wages in terms of tasks and time spent on tasks (in seconds). To do so, we use the fact that across all treatments and scenarios, the average reservation wage for 15 additional tasks is never higher than $\$2.99$, meaning that (on average) people are willing to do an additional task for $\$2.99/15 \approx \$0.20$, so that $\$0.82$ is equivalent to about 4 tasks. This in turn is equivalent to 3 minutes (188 seconds), given that the average time spent per task is $46$ seconds.

\import{./data/}{table_cost_of_bracketing.tex}

We also report the change in reservation wages by gender based on \cite{koch2019correlates}, who state that ``[w]omen seem to be more prone to narrow bracketing than men''. The estimated changes range from $\$0.14$ (Scenario 2) to $\$0.41$ (Scenario 1) in the pooled data, from $\$0.13$ to $\$0.57$ for female and from $\$0.16$ to $\$0.30$ for male participants. Thus we also find that women have larger changes due to bracketing. Note however that under full narrow bracketing, since the reservation wage in BOTH should equal that in MONEY/LOW, we would expect that $\Delta := m_{NONE} - m_{BOTH} = m_{NONE} - m_{MONEY/LOW}$. Since $m_{NONE}$ measures the marginal disutility when the baseline is 15 tasks higher than for $m_{MONEY/LOW}$, this is a measure of the convexity of disutility, and $\Delta$ is predicted to be larger the more convex the preferences are. We therefore report $\hat{\Delta} := m_{NONE} - m_{MONEY/LOW}$ in Table \ref{tab:cost_of_bracketing}.

This shows that the results are in line with both men and women bracketing narrowly, yet women having larger changes in reservation wages due to more convex disutility from work. For women, the marginal disutility of doing 15 sequences on top of 0 sequences (elicited in Scenario 1 of MONEY/LOW) is $\$0.91$ higher than their marginal disutility of doing 15 sequences on top of 15 sequences (as elicited in Scenario 1 of BOTH). For men, this figure stands only at $\$0.38$. For both genders, we can reject broad bracketing, but not narrow bracketing (see Appendix \ref{appendix:wilcox-and-gender} for details). This suggests that some of the gender differences in \cite{koch2019correlates} may be due to gender differences in preferences.


\subsection{Bracketing in Work and Money Dimensions Separately}\label{app:bracketing-work-and-money}

Our results show that people do not bracket broadly. It is still possible that this is only due to a failure to bracket broadly their endowment of money, even though they broadly bracket their work endowment. We therefore ran an additional treatment, MONEY, in which there was only an endowment of MONEY, but no endowment of work. Thus, if participants in treatment BOTH bracketed work broadly, they should behave identically to participants in MONEY in both scenarios. We did not preregister for this treatment and we ran it after having started our main treatments. For this reason, the participants in MONEY are not balanced by session against the other treatments. See Appendix \ref{appendix:implementation-details} for details.

Formally, we have:

\begin{hypothesis}[Broad Bracketing of Work]\label{hyp:broad-work}
    Behavior is consistent with broadly bracketing work if $m_{MONEY} = m_{BOTH}$ in every Scenario. 
\end{hypothesis}

\begin{hypothesis}[Broad Bracketing of Money]\label{hyp:broad-money}
    Behavior is consistent with broadly bracketing money if $m_{MONEY} = m_{NONE}$ in every Scenario.
\end{hypothesis}

Table \ref{tab:cost_of_work_bracketing} reports the differences in means between treatments MONEY and BOTH. The results show a sizeable and statistically significant difference in scenario 1 (p-value: $0.002$), and a smaller and statistically insignificant difference for Scenario 2, similar to the main results. We thus reject that participants broadly bracket work.\footnote{This is the only result that is not statistically significant when we instead use a linear regression that averages differences in reservation wages across scenarios -- see Appendix \ref{appendix:linear-regression}. The reason is clear from \ref{tab:cost_of_work_bracketing}: in Scenario 1, the difference between BOTH and MONEY is $2.07-2.50 = -0.43$, while it is $2.70-2.43 = 0.27$ in Scenario 2, which averages out to a difference of $(-0.43 + 0.27)/2 = 0.08$ that is indistinguishable from 0. Since bracketing predicts identical behavior for each Scenario, averaging across scenarios as the linear regression does is not the appropriate test and overly conservative -- both for testing broad \emph{and} narrow bracketing.}

\begin{result}
  We reject Hypothesis \ref{hyp:broad-work} that individuals bracket the work dimension broadly.
\end{result}

Regarding gender differences, we see that men's choices are consistent with broad bracketing of work, while women's choices are not. However, men's choices are also consistent with full narrow bracketing, because men's choices are consistent with linear costs, while women's choices aren't. Thus the identification assumption 
fails for men and holds for women. Thus while it might be justified to claim that women incur larger costs than men from bracketing work, it would be wrong to claim that they are bracketing work more narrowly.

\import{./data/}{table_cost_of_work_bracketing.tex}

Table \ref{tab:cost_of_money_bracketing} similarly shows that people do not broadly bracket money, and that both men and women bracket it narrowly.

\import{./data/}{table_cost_of_money_bracketing.tex}

\begin{result}
  We reject the Hypothesis \ref{hyp:broad-money} that individuals bracket the money dimension broadly.
\end{result}

\begin{figure}[H]
  \caption{A bar plot of the reservation wages for treatments NONE, BOTH, and MONEY by scenario.}
  \includegraphics[width=0.95\textwidth]{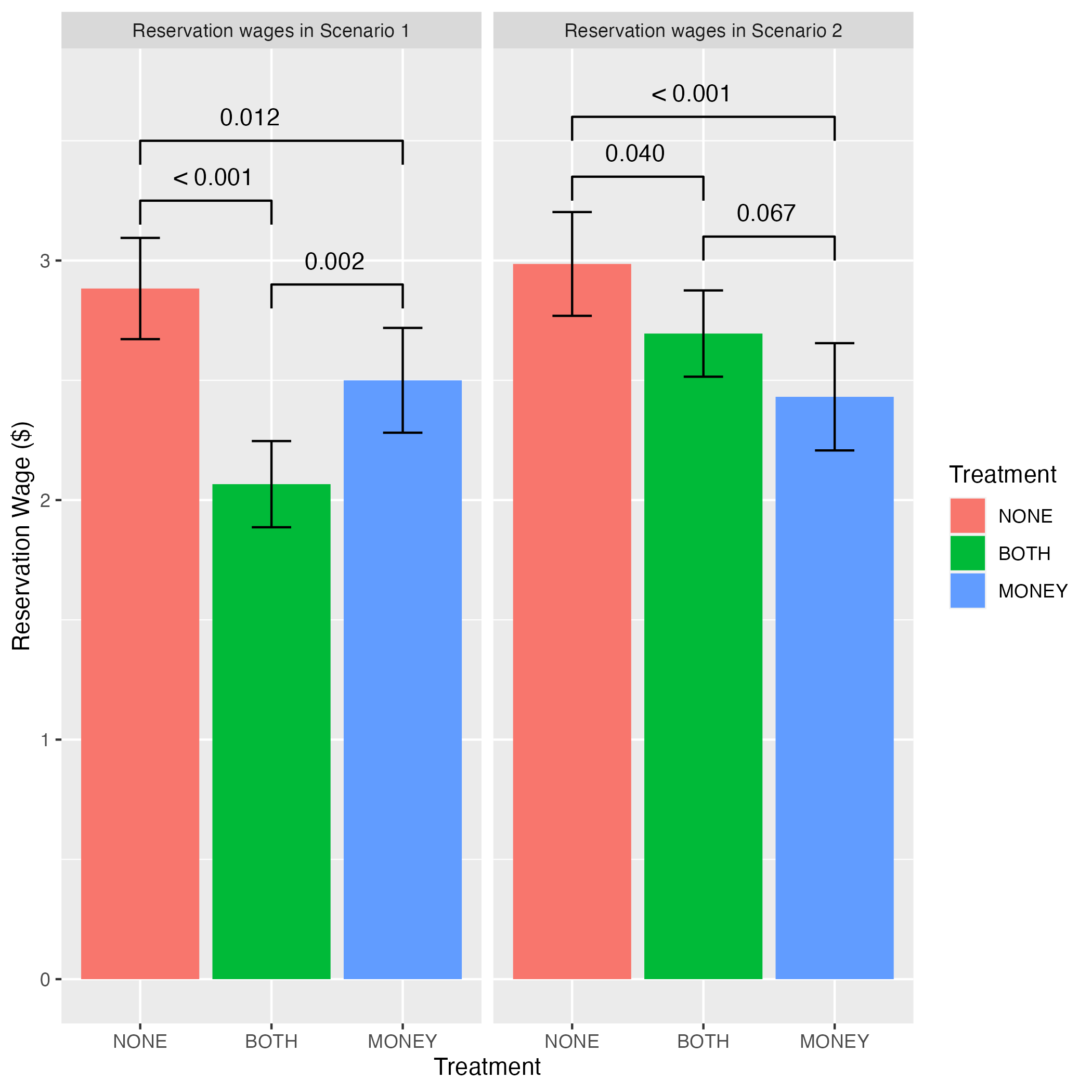}\label{fig:bar-plot-money}
\end{figure}

Figure \ref{fig:bar-plot-money} provides a bar plot of all the means by scenario.

%% file: data/mwu_all_tables.tex
\begin{table}[H]
\centering
\caption{\label{tab:mwu_all_tables}Between-treatment p-values for main treatments based on two-sided Wilcoxon rank-sum tests, treating each individual in each scenario as a single independent observation. The first two columns are for pooled data, the next two for data restricted to female participants, the final two for data restricted to male participants.}
\centering
\fontsize{12}{14}\selectfont
\begin{tabular}[t]{lcccccc}
\toprule
\multicolumn{3}{c}{Pooled} & \multicolumn{2}{c}{Female} & \multicolumn{2}{c}{Male} \\
\cmidrule(l{3pt}r{3pt}){1-3} \cmidrule(l{3pt}r{3pt}){4-5} \cmidrule(l{3pt}r{3pt}){6-7}
Treatments & NONE & BOTH & NONE/F & BOTH/F & NONE/M & BOTH/M\\
\midrule
\addlinespace[0.3em]
\multicolumn{7}{l}{\textbf{Scenario 1}}\\
\hspace{1em}BOTH & $< 0.001$ &  & $< 0.001$ &  & $0.002$ & \\
\hspace{1em}MONEY/LOW & $< 0.001$ & $0.117$ & $0.001$ & $0.445$ & $0.052$ & $0.300$\\
\addlinespace[0.3em]
\multicolumn{7}{l}{\textbf{Scenario 2}}\\
\hspace{1em}BOTH & $0.023$ &  & $0.277$ &  & $0.038$ & \\
\hspace{1em}MONEY/LOW & $0.108$ & $0.704$ & $0.357$ & $0.984$ & $0.121$ & $0.809$\\
\bottomrule
\end{tabular}
\end{table}

%% file: data/t_test_all_tables.tex
\begin{table}[H]
\centering
\caption{\label{tab:t_test_all_tables}Between-treatment p-values for main treatments based on two-sided t-tests by scenario, treating each individual in each scenario as a single independent observation. The first two columns are for pooled data, the next two for data restricted to female participants, the final two for data restricted to male participants.}
\centering
\fontsize{12}{14}\selectfont
\begin{tabular}[t]{lcccccc}
\toprule
\multicolumn{3}{c}{Pooled} & \multicolumn{2}{c}{Female} & \multicolumn{2}{c}{Male} \\
\cmidrule(l{3pt}r{3pt}){1-3} \cmidrule(l{3pt}r{3pt}){4-5} \cmidrule(l{3pt}r{3pt}){6-7}
Treatments & NONE & BOTH & NONE/F & BOTH/F & NONE/M & BOTH/M\\
\midrule
\addlinespace[0.3em]
\multicolumn{7}{l}{\textbf{Scenario 1}}\\
\hspace{1em}BOTH & $< 0.001$ &  & $< 0.001$ &  & $0.001$ & \\
\hspace{1em}MONEY/LOW & $< 0.001$ & $0.106$ & $< 0.001$ & $0.342$ & $0.055$ & $0.241$\\
\addlinespace[0.3em]
\multicolumn{7}{l}{\textbf{Scenario 2}}\\
\hspace{1em}BOTH & $0.040$ &  & $0.264$ &  & $0.072$ & \\
\hspace{1em}MONEY/LOW & $0.120$ & $0.753$ & $0.383$ & $0.909$ & $0.133$ & $0.886$\\
\bottomrule
\end{tabular}
\end{table}

%% file: data/session_summary.tex
\begin{table}[!h]
\centering
\caption{\label{tab:session_summary}Participant numbers by sessions and treatments}
\centering
\fontsize{12}{14}\selectfont
\begin{tabular}[t]{rrrrrr}
\toprule
Session ID & Session Date & NONE & BOTH & MONEY/LOW & MONEY\\
\midrule
1 & 2019-12-18 & 9 & 9 & 8 & 0\\
2 & 2019-12-19 & 11 & 12 & 13 & 0\\
3 & 2019-12-19 & 39 & 38 & 40 & 0\\
4 & 2019-12-19 & 18 & 20 & 19 & 0\\
5 & 2019-12-20 & 68 & 0 & 0 & 0\\
\addlinespace
6 & 2019-12-21 & 0 & 68 & 0 & 0\\
7 & 2019-12-23 & 0 & 0 & 61 & 0\\
8 & 2019-12-30 & 23 & 24 & 23 & 0\\
9 & 2020-01-21 & 15 & 13 & 15 & 43\\
10 & 2020-01-22 & 5 & 4 & 5 & 14\\
\addlinespace
11 & 2020-01-28 & 12 & 13 & 12 & 36\\
12 & 2020-01-30 & 0 & 0 & 0 & 68\\
13 & 2020-02-04 & 0 & 0 & 0 & 52\\
14 & 2020-02-05 & 0 & 38 & 0 & 0\\
15 & 2020-02-06 & 0 & 81 & 0 & 0\\
\midrule\\
\addlinespace
 &  & 200 & 320 & 196 & 213\\
\bottomrule
\end{tabular}
\end{table}

%% file: data/attrition_statistics.tex
\begin{table}
\centering
\caption{\label{tab:attrition}Attrition in \% by a given stage}
\centering
\fontsize{12}{14}\selectfont
\begin{tabular}[t]{lrrrrr}
\toprule
Treatments & Practice & Choice 1 & Answer 1 & Learn Tasks & End\\
\midrule
\addlinespace[0.3em]
\multicolumn{6}{l}{\textbf{Main}}\\
\hspace{1em}\cellcolor{gray!10}{NONE} & \cellcolor{gray!10}{11\%} & \cellcolor{gray!10}{11\%} & \cellcolor{gray!10}{11\%} & \cellcolor{gray!10}{12\%} & \cellcolor{gray!10}{18\%}\\
\hspace{1em}BOTH & 11\% & 11\% & 11\% & 12\% & 20\%\\
\hspace{1em}\cellcolor{gray!10}{MONEY/LOW} & \cellcolor{gray!10}{9\%} & \cellcolor{gray!10}{10\%} & \cellcolor{gray!10}{10\%} & \cellcolor{gray!10}{11\%} & \cellcolor{gray!10}{13\%}\\
\hspace{1em}MONEY & 8\% & 8\% & 8\% & 8\% & 16\%\\
\addlinespace[0.3em]
\multicolumn{6}{l}{\textbf{Follow Up}}\\
\hspace{1em}\cellcolor{gray!10}{BEFORE} & \cellcolor{gray!10}{7\%} & \cellcolor{gray!10}{7\%} & \cellcolor{gray!10}{7\%} & \cellcolor{gray!10}{7\%} & \cellcolor{gray!10}{18\%}\\
\hspace{1em}AFTER & 7\% & 7\% & 7\% & 7\% & 14\%\\
\bottomrule
\end{tabular}
\end{table}

%% file: data/t_test_balanced.tex
\begin{table}
\centering
\caption{\label{tab:t_test_balanced} Between-treatment p-values for NONE, BOTH, and MONEY/LOW treatments based on two-sided t-test, treating each individual in each scenario as a single independent observation. Restricted to sessions in which these three treatments were balanced.}
\centering
\fontsize{12}{14}\selectfont
\begin{tabular}[t]{lcc}
\toprule
Treatments & NONE & BOTH\\
\midrule
\addlinespace[0.3em]
\multicolumn{3}{l}{\textbf{Scenario 1}}\\
\hspace{1em}BOTH & $< 0.001$ & \\
\hspace{1em}MONEY/LOW & $0.003$ & $0.316$\\
\addlinespace[0.3em]
\multicolumn{3}{l}{\textbf{Scenario 2}}\\
\hspace{1em}BOTH & $0.245$ & \\
\hspace{1em}MONEY/LOW & $0.222$ & $0.936$\\
\bottomrule
\end{tabular}
\end{table}

%% file: data/mwu_balanced.tex
\begin{table}
\centering
\caption{\label{tab:mwu_balanced}Between-treatment p-values for NONE, BOTH, and MONEY/LOW treatments based on two-sided Wilcoxon rank-sum tests, treating each individual in each scenario as a single independent observation. Restricted to sessions in which these three treatments were balanced.}
\centering
\fontsize{12}{14}\selectfont
\begin{tabular}[t]{lcc}
\toprule
Treatments & NONE & BOTH\\
\midrule
\addlinespace[0.3em]
\multicolumn{3}{l}{\textbf{Scenario 1}}\\
\hspace{1em}BOTH & $< 0.001$ & \\
\hspace{1em}MONEY/LOW & $0.004$ & $0.369$\\
\addlinespace[0.3em]
\multicolumn{3}{l}{\textbf{Scenario 2}}\\
\hspace{1em}BOTH & $0.270$ & \\
\hspace{1em}MONEY/LOW & $0.284$ & $0.941$\\
\bottomrule
\end{tabular}
\end{table}

%% file: data/t_test_presentation_before_page.tex
\begin{table}
\centering
\caption{\label{tab:t_test_test_before_page}Between-treatment p-values for main treatments based on two-sided t-test, treating each individual in each scenario as a single independent observation. Restricted to those sessions of BOTH where baseline is revealed right before the first choice page.}
\centering
\fontsize{12}{14}\selectfont
\begin{tabular}[t]{lcc}
\toprule
Treatments & NONE & BOTH\\
\midrule
\addlinespace[0.3em]
\multicolumn{3}{l}{\textbf{Scenario 1}}\\
\hspace{1em}BOTH & $< 0.001$ & \\
\hspace{1em}MONEY/LOW & $< 0.001$ & $0.243$\\
\addlinespace[0.3em]
\multicolumn{3}{l}{\textbf{Scenario 2}}\\
\hspace{1em}BOTH & $0.400$ & \\
\hspace{1em}MONEY/LOW & $0.120$ & $0.490$\\
\bottomrule
\end{tabular}
\end{table}

%% file: data/t_test_narrow_v_narrow.tex
\begin{table}
\centering
\caption{\label{tab:t_test_narrow_v_narrow}Between-treatment p-values for BOTH when information on baseline is presented for the first time right before the first choice or exactly on the first choice page. Based on two-sided Wilcoxon rank-sum tests, treating each individual in each scenario as a single independent observation.}
\centering
\fontsize{12}{14}\selectfont
\begin{tabular}[t]{lr}
\toprule
Scenarios & t-test\\
\midrule
Scenario 1 & $0.551$\\
Scenario 2 & $0.001$\\
\bottomrule
\end{tabular}
\end{table}

%% file: data/rw_change_between_scenarios.tex
\begin{table}[H]
\centering
\caption{\label{tab:rw_change_between_scenarios}Average individual-level change in reservation wage, conditional on whether the jump was up or down.}
\centering
\fontsize{12}{14}\selectfont
\begin{tabular}[t]{lrr}
\toprule
Treatment & Down & Up\\
\midrule
\cellcolor{gray!10}{NONE} & \cellcolor{gray!10}{-1.10} & \cellcolor{gray!10}{0.83}\\
BOTH & -0.83 & 1.07\\
\cellcolor{gray!10}{MONEY/LOW} & \cellcolor{gray!10}{-0.69} & \cellcolor{gray!10}{0.94}\\
MONEY & -1.10 & 0.93\\
\bottomrule
\end{tabular}
\end{table}

%% file: data/rw_direction_change_between_scenarios.tex
\begin{table}
\centering
\caption{\label{tab:rw_direction_between_scenarios}Frequencies (in \%) of individuals who switch up, switch down, or stay at the same reservation wage from scenario 1 to scenario 2. The final column reports how many more people switch up rather than down.}
\centering
\fontsize{12}{14}\selectfont
\begin{tabular}[t]{lrrrrr}
\toprule
Treatment & Down & Stay & Up & Drop out & Up - Down\\
\midrule
\cellcolor{gray!10}{NONE} & \cellcolor{gray!10}{16} & \cellcolor{gray!10}{44} & \cellcolor{gray!10}{34} & \cellcolor{gray!10}{6} & \cellcolor{gray!10}{18}\\
BOTH & 6 & 27 & 60 & 8 & 54\\
\cellcolor{gray!10}{MONEY/LOW} & \cellcolor{gray!10}{10} & \cellcolor{gray!10}{38} & \cellcolor{gray!10}{49} & \cellcolor{gray!10}{4} & \cellcolor{gray!10}{39}\\
MONEY & 30 & 42 & 26 & 2 & -4\\
\bottomrule
\end{tabular}
\end{table}

%% file: data/linear_regression_pooled_and_by_scenario.tex
\begin{table}[!htbp] \centering 
  \caption{Linear Regressions of reservation wages by treatment averaged across scenarios. With and without clustered standard errors by participant. The default treatment is NONE, so that broad bracketing predicts a null estimate for the fixed effect for BOTH (rejected), and narrow bracketing predicts equal fixed effects for BOTH and MONEY/FALSE (not rejected). Broad bracketing of money (but not necessarily work) predicts a null estimate for MONEY (rejected); broad bracketing of work (but not necessarily money) predicts equal fixed effects for BOTH and MONEY (not rejected). The last-mentioned test is not rejected because the difference is positive in Scenario 1 and negative in Scenario 2, which is average in this regression. Since bracketing makes predictions that should hold for every scenario, the test in our main text is the better test (accounting for double-testing).} 
  \label{tab:linear-regressions} 
\begin{tabular}{@{\extracolsep{5pt}}lcc} 
\\[-1.8ex]\hline 
\hline \\[-1.8ex] 
 & \multicolumn{2}{c}{\textit{Dependent variable:}} \\ 
\cline{2-3} 
\\[-1.8ex] & \multicolumn{2}{c}{Reservation wage} \\ 
 & No clustering & Clustering by participant \\ 
\\[-1.8ex] & (1) & (2)\\ 
\hline \\[-1.8ex] 
 scenarioScenario2 & 0.31$^{}$ & 0.31$^{}$ \\ 
  & (0.07) & (0.04) \\ 
  & & \\ 
 treatmentBOTH & $-$0.55$^{}$ & $-$0.55$^{}$ \\ 
  & (0.10) & (0.13) \\ 
  & & \\ 
 treatmentMONEY & $-$0.47$^{}$ & $-$0.47$^{}$ \\ 
  & (0.11) & (0.14) \\ 
  & & \\ 
 treatmentMONEY/LOW & $-$0.41$^{}$ & $-$0.41$^{}$ \\ 
  & (0.11) & (0.15) \\ 
  & & \\ 
 Constant & 2.78$^{}$ & 2.78$^{}$ \\ 
  & (0.09) & (0.10) \\ 
  & & \\ 
\hline \\[-1.8ex] 
Observations & 1,231 & 1,231 \\ 
R$^{2}$ & 0.04 & 0.04 \\ 
F Statistic (df = 4; 1226) & 12.62$^{}$ & 12.62$^{}$ \\ 
\hline 
\hline \\[-1.8ex] 
\textit{Note:}  & \multicolumn{2}{r}{Standard errors in parentheses.} \\ 
\end{tabular} 
\end{table}

%% file: data/means_work_bracketing.tex
\begin{table}
\centering
\caption{\label{tab:means_work_bracketing}Mean comparisons between NONE, MONEY, and BOTH}
\centering
\fontsize{12}{14}\selectfont
\begin{tabular}[t]{llrrr}
\toprule
Scenario & Treatment & Mean & N & Std. Err.\\
\midrule
Scenario1 & BOTH & 2.07 & 208 & 0.09\\
Scenario2 & BOTH & 2.70 & 206 & 0.09\\
Scenario1 & NONE & 2.88 & 137 & 0.11\\
Scenario2 & NONE & 2.99 & 141 & 0.11\\
Scenario1 & MONEY & 2.50 & 137 & 0.11\\
\addlinespace
Scenario2 & MONEY & 2.43 & 138 & 0.11\\
\bottomrule
\end{tabular}
\end{table}

%% file: data/means_data.tex
\begin{table}[!h]
\centering
\caption{\label{tab:means_data}Means of main treatments by scenario}
\centering
\fontsize{12}{14}\selectfont
\begin{tabular}[t]{lrrrr}
\toprule
Treatment & Res. Wage & Std Err & \% upper bound & N\\
\midrule
\addlinespace[0.3em]
\multicolumn{5}{l}{\textbf{Scenario 1}}\\
\hspace{1em}NONE & 2.88 & 0.11 & 24\% & 137\\
\hspace{1em}BOTH & 2.07 & 0.09 & 12\% & 207\\
\hspace{1em}MONEY/LOW & 2.31 & 0.12 & 18\% & 131\\
\addlinespace[0.3em]
\multicolumn{5}{l}{\textbf{Scenario 2}}\\
\hspace{1em}NONE & 2.99 & 0.11 & 33\% & 141\\
\hspace{1em}BOTH & 2.70 & 0.09 & 20\% & 205\\
\hspace{1em}MONEY/LOW & 2.74 & 0.11 & 25\% & 135\\
\bottomrule
\end{tabular}
\end{table}

%% file: data/table_cost_of_bracketing.tex
\begin{table}
\centering
\caption{\label{tab:cost_of_bracketing}The table reports $\Delta$, the change in reservation wages between NONE and BOTH. The highest average reservation wage for 15 more tasks is 2.99 across all treatments and scenarios, so that $2.99/15 \approx 0.20$ is an upper bound for the average cost per task. Using this, we can convert $\Delta$ into task-equivalents by $\Delta / 0.20$, and the cost in time-equivalents (in seconds) by $(\Delta / 0.20) / 46$, since the average time taken for a task is 46 seconds. \emph{$\hat{\Delta}$} stands for $m_{NONE} - m_{MONEY/LOW}$: the change in the marginal disutility of doing 15 extra tasks on top of a low vs on top of a high baseline. Under full narrow bracketing, \emph{$\hat{\Delta}$} and $\Delta$ should be equal.}
\centering
\fontsize{12}{14}\selectfont
\begin{tabular}[t]{llrrrrr}
\toprule
Scenario & Gender & $\Delta$ & Std.Err. & $\hat{\Delta}$ & Task equivalent & Time equivalent (in secs)\\
\midrule
\addlinespace[0.3em]
\multicolumn{7}{l}{\textbf{Pooled}}\\
\hspace{1em}Scenario1 & Pooled & 0.82 & 0.14 & 0.58 & 4.10 & 188\\
\hspace{1em}Scenario2 & Pooled & 0.29 & 0.14 & 0.25 & 1.46 & 67\\
\addlinespace[0.3em]
\multicolumn{7}{l}{\textbf{Female}}\\
\hspace{1em}Scenario1 & Female & 1.14 & 0.21 & 0.91 & 5.71 & 262\\
\hspace{1em}Scenario2 & Female & 0.26 & 0.23 & 0.23 & 1.28 & 59\\
\addlinespace[0.3em]
\multicolumn{7}{l}{\textbf{Male}}\\
\hspace{1em}Scenario1 & Male & 0.60 & 0.18 & 0.38 & 3.01 & 138\\
\hspace{1em}Scenario2 & Male & 0.32 & 0.18 & 0.30 & 1.61 & 74\\
\bottomrule
\end{tabular}
\end{table}

%% file: data/table_cost_of_work_bracketing.tex
\begin{table}
\centering
\caption{\label{tab:cost_of_work_bracketing}The table reports $\Delta$, the change in reservation wages between MONEY and BOTH. Under broad bracketing of work, $\Delta$ should be $0$, see Hypothesis \ref{hyp:broad-work}. \emph{$\hat{\Delta}$} stands for $m_{MONEY} - m_{MONEY/LOW}$. When $\hat{\Delta} = 0$, we cannot identify full narrow bracketing from broad bracketing of work.}
\centering
\fontsize{12}{14}\selectfont
\begin{tabular}[t]{llrrr}
\toprule
Scenario & Gender & $\Delta$ & Std. Err. & $\hat{\Delta}$\\
\midrule
\addlinespace[0.3em]
\multicolumn{5}{l}{\textbf{Pooled}}\\
\hspace{1em}Scenario1 & Pooled & 0.43 & 0.14 & 0.19\\
\hspace{1em}Scenario2 & Pooled & -0.26 & 0.14 & -0.31\\
\addlinespace[0.3em]
\multicolumn{5}{l}{\textbf{Female}}\\
\hspace{1em}Scenario1 & Female & 0.65 & 0.22 & 0.42\\
\hspace{1em}Scenario2 & Female & -0.26 & 0.24 & -0.29\\
\addlinespace[0.3em]
\multicolumn{5}{l}{\textbf{Male}}\\
\hspace{1em}Scenario1 & Male & 0.28 & 0.19 & 0.06\\
\hspace{1em}Scenario2 & Male & -0.28 & 0.18 & -0.31\\
\bottomrule
\end{tabular}
\end{table}

%% file: data/table_cost_of_money_bracketing.tex
\begin{table}
\centering
\caption{\label{tab:cost_of_money_bracketing}The table reports $\Delta$, the change in reservation wages between MONEY and NONE. Broad bracketing of money requires $\Delta = 0$. \emph{$\hat{\Delta}$} stands for $m_{BOTH} - m_{MONEY/LOW}$. Full narrow bracketing requires $\hat{\Delta} = 0$.}
\centering
\fontsize{12}{14}\selectfont
\begin{tabular}[t]{llrrr}
\toprule
Scenario & Gender & $\Delta$ & Std. Err. & $\hat{\Delta}$\\
\midrule
\addlinespace[0.3em]
\multicolumn{5}{l}{\textbf{Pooled}}\\
\hspace{1em}Scenario1 & Pooled & -0.38 & 0.15 & -0.24\\
\hspace{1em}Scenario2 & Pooled & -0.55 & 0.16 & -0.05\\
\addlinespace[0.3em]
\multicolumn{5}{l}{\textbf{Female}}\\
\hspace{1em}Scenario1 & Female & -0.49 & 0.24 & -0.23\\
\hspace{1em}Scenario2 & Female & -0.51 & 0.26 & -0.03\\
\addlinespace[0.3em]
\multicolumn{5}{l}{\textbf{Male}}\\
\hspace{1em}Scenario1 & Male & -0.32 & 0.20 & -0.22\\
\hspace{1em}Scenario2 & Male & -0.60 & 0.20 & -0.03\\
\bottomrule
\end{tabular}
\end{table}

%% file: appendix-instructions-study2.tex
\section{Appendix: Study 2 Instructions}\label{appendix:study2-instructions}
\subsection{Main Task Instructions}

\textbf{THE MAIN STUDY}

\textbf{We now describe the main study.}

\textbf{Guaranteed Participation Bonus:}

\begin{itemize}
    \item For completing the study, you will receive a guaranteed payment of £3.00.
\end{itemize}

\textbf{Additional Bonus}
\begin{itemize}
    \item You have the chance to earn an additional bonus based on your decisions and performance. 
    \item Specifically, you can earn this bonus by successfully decoding a certain number of sequences.
\end{itemize}

\textbf{Study Structure:}
\begin{itemize}
    \item During the study, you will encounter 35 choice pages. Each page presents a different scenario with various sets of sequences for you to decode.
    \item Your choice is to select your preferred option from these scenarios on each choice page.
\end{itemize}

\textbf{Bonus Determination:}

\begin{itemize}
    \item After you make your decisions, the computer will randomly select one of the 35 scenarios as the basis for your bonus payment and workload.
    \item This means that the number of sequences you need to decode and any additional bonus you earn will depend on the scenario selected by the computer.
\end{itemize}

\textbf{Further Instructions:}

\begin{itemize}
    \item Detailed explanations of how your decisions influence the sequences and bonuses will be provided on the next page.
\end{itemize}

\subsection{Study Description - Part 1}

In each decision screen, you will be presented with one of three different types of SCENARIOS.

\textbf{Type 1 SCENARIOS}

The first type is a choice between two or more alternatives, each involving different bonuses and sequences to decode.

Example (DOES NOT COUNT):

\textit{Choose your preferred option:}

\begin{itemize}
    \item Option A: Decode 5 sequences for £0.30
    \item Option B: Decode 1 sequence for £0.00
    \item Option C: Decode 2 sequences for £0.10
\end{itemize}

If, for example, you select option C, you will decode 2 sequences and receive £0.10.

\textbf{Type 2 SCENARIOS}

The second type is a choice of how many sequences you are willing to decode if they are paid a certain sum each.

Example (DOES NOT COUNT):

\textit{You can decode up to 10 sequences, and you will earn £0.05 per sequence decoded correctly.}

\textbf{How many sequences would you like to decode?}
\\

If, for example, you choose that you are willing to decode 5 sequences for £0.05, that means that you are not willing to decode more than 5 sequences for such payment.
\\

\textbf{Type 3 SCENARIOS}

In these scenarios, the options will involve a certain probability of decoding sequences.
\\

Example (DOES NOT COUNT):
\\
\textit{Choose your preferred option:}
\begin{itemize}
    \item Option A: Receive £1.00 for sure. And if the six-sided virtual die rolls 1 decode 5 sequences; otherwise, decode 0 sequences
    \item Option B: Decode 1 sequence for £0.00
\end{itemize}
\vspace{0.5em}
Probability-based selection process: \\

If, for example, you select Option A, you will receive £1.00 for sure. Additionally, you will either decode 5 sequences (if the virtual die rolls a 1) or 0 sequences (otherwise).

\subsection{Study Description - Part 2}

Some Scenarios will have two choices like the one presented earlier to be taken at the same time. In that case, the output of bonuses and sequences to decode of both choices will matter.
\\

Example (DOES NOT COUNT):
\\
In this Scenario you have to make two decisions. 

\textbf{Decision 1}
\\
\textit{Choose your preferred option:}
\begin{itemize}
    \item Option A: Decode 5 sequences for £0.30
    \item Option B: Decode 1 sequence for £0.00
    \item Option C: Decode 2 sequence for £0.10
\end{itemize}
\vspace{0.5em}
Before answering, read the next decision
\\
\textbf{Decision 2}
\\
\textit{Choose your preferred option:}
\begin{itemize}
    \item Option D: Decode 10 sequences for £1.00
    \item Option E: Decode 1 sequence for £0.00
\end{itemize}
\vspace{0.5em}
Suppose that in the example above you choose Option A in Choice 1 and Option D in Choice 2.
\\
If the computer selects this Scenario, then you will be asked to decode 15 (5+10) sequences and you will receive an additional bonus of £1.30 (£0.30+£1.00).
\\

\textbf{Selection Process:} The computer will randomly select one scenario at the end of all the choices.
\\
Based on the choices you made in this randomly selected scenario, you will be asked to decode the sequences and receive the payment associated with your chosen option.
\\

Suppose that in the example above you choose Option A in Choice 1 and Option D in Choice 2.

If the computer selects this Scenario, then you will be asked to decode 15 (5+10) sequences and you will receive an additional £1.30 (£0.30+£1.00).

\subsection{Study Description - Summary}

\textbf{TO SUMMARIZE}

\begin{itemize}
    \item You will make decisions in 35 Scenarios in total.
    \item The computer will randomly select one of these Scenarios.
    \item For this selected Scenario you will be asked to decode the sequences and receive a bonus depending on the choice(s) you made in that Scenario.
\end{itemize}

\subsection{Comprehension Questions}\label{appendix:controlquestions}

\textbf{QUIZ}
\\
\textbf{Before proceeding you have to answer to the following questions. If you fail to answer twice the study will finish and you will not receive any payment.}

\textbf{Which of the following is true?}
\\
\begin{itemize}
    \item Only the decisions made in one of the Scenarios will be relevant for the payment of the bonus.
    \item The bonus sums up the decisions made in all Scenarios.
\end{itemize}
\vspace{0.5em}
\textbf{Which of the following is true?} \\
When more decisions are shown under the same Scenario page..
\vspace{0.5em}
\begin{itemize}
    \item ..only one of them will be relevant for the payment of the bonus.
    \item ..all the decisions will be relevant for the payment of the bonus.
\end{itemize}
\vspace{0.5em}

\subsection{Main Task - list of Scenarios presented individually}
\vspace{0.5cm}
SCENARIO 1 Choose your preferred option:

\begin{itemize}
    \item OPTION A: Decode 0 sequences for £0.25
    \item OPTION B: Decode 10 sequences for £0.45
    \item OPTION C: Decode 10 sequences for £0.40
\end{itemize}
\vspace{0.5cm}

SCENARIO 2
Choose your preferred option:

\begin{itemize}
    \item OPTION A: Decode 5 sequences for £0.45
    \item OPTION B: Decode 12 sequences for £0.80
\end{itemize}

\vspace{0.5cm}

SCENARIO 3
Choose your preferred option:
\begin{itemize}
    \item OPTION A: Decode 7 sequences for £0.40
    \item OPTION B: Decode 13 sequences for £0.45
    \item OPTION C: Decode 15 sequences for £0.70
\end{itemize}

\vspace{0.5cm}

SCENARIO 4
Choose your preferred option:
\begin{itemize}
    \item OPTION A: Decode 4 sequences for £0.40
    \item OPTION B: Decode 14 sequences for £0.79
\end{itemize}

\vspace{0.5cm}

SCENARIO 5
Choose your preferred option:
\begin{itemize}
    \item OPTION A: Decode 5 sequences for £0.45
    \item OPTION B: Decode 12 sequences for £0.75
    \item OPTION C: Decode 12 sequences for £0.70
\end{itemize}

\vspace{0.5cm}

SCENARIO 6
Choose your preferred option:
\begin{itemize}
    \item OPTION A: Decode 5 sequences for £0.35
    \item OPTION B: Decode 16 sequences for £0.90
\end{itemize}

\vspace{0.5cm}

SCENARIO 7
Choose your preferred option:
\begin{itemize}
    \item OPTION A: Decode 5 sequences for £0.40
    \item OPTION B: Decode 12 sequences for £0.75
    \item OPTION C: Decode 10 sequences for £0.50
\end{itemize}

\vspace{0.5cm}

SCENARIO 8
Choose your preferred option:
\begin{itemize}
    \item OPTION A: Decode 10 sequences for £0.45
    \item OPTION B: Decode 14 sequences for £0.75
    \item OPTION C: Decode 4 sequences for £0.40
\end{itemize}

\vspace{0.5cm}

SCENARIO 9
Choose your preferred option:
\begin{itemize}
    \item OPTION A: Decode 15 sequences for £0.75
    \item OPTION B: Decode 7 sequences for £0.40
\end{itemize}

\vspace{0.5cm}

SCENARIO 10
Choose your preferred option:
\begin{itemize}
    \item OPTION A: Decode 7 sequences for £0.40
    \item OPTION B: Decode 9 sequences for £0.45
    \item OPTION C: Decode 15 sequences for £0.70
\end{itemize}

\vspace{0.5cm}

SCENARIO 11
Choose your preferred option:
\begin{itemize}
    \item OPTION A: Decode 0 sequences for £0.25
    \item OPTION B: Decode 10 sequences for £0.50
\end{itemize}

\vspace{0.5cm}

SCENARIO 12
Choose your preferred option:
\begin{itemize}
    \item OPTION A: Decode 5 sequences for £0.35
    \item OPTION B: Decode 16 sequences for £0.85
    \item OPTION C: Decode 14 sequences for £0.45
\end{itemize}

\vspace{0.5cm}

SCENARIO 13
Description

You can decode up to 25 sequences, and you will earn £0.12 per sequence decoded correctly.

How many sequences would you like to decode?

\vspace{0.5cm}
SCENARIO 14
Description

You can decode up to 25 sequences, and you will earn £0.08 per sequence decoded correctly.

How many sequences would you like to decode?

\vspace{0.5cm}
SCENARIO 15
Description

You can decode up to 25 sequences, and you will earn £0.05 per sequence decoded correctly.

How many sequences would you like to decode?

\vspace{0.5cm}
SCENARIO 16
Description

You can decode up to 25 sequences, and you will earn £0.03 per sequence decoded correctly.

How many sequences would you like to decode?

\vspace{0.5cm}
SCENARIO 17
Description

You can decode up to 25 sequences, and you will earn £0.02 per sequence decoded correctly.

How many sequences would you like to decode?

\vspace{0.5cm}
SCENARIO 18
Choose your preferred option:
\begin{itemize}
    \item OPTION A: Receive £0.30 for sure. And if the virtual six-sided die rolls 6, decode 20 sequences; otherwise, decode 0 sequences
    \item OPTION B: Decode 0 sequences for £0.15
\end{itemize}

\vspace{0.5cm}

SCENARIO 19
Choose your preferred option:

\begin{itemize}
    \item OPTION A: Receive £0.40 for sure. And if the virtual six-sided die rolls 4 or 5, decode 20 sequences; otherwise, decode 0 sequences
\item OPTION B: Receive £0.30 for sure. And if the virtual six-sided die rolls 5 decode 20 sequences; otherwise, decode 0 sequences
\end{itemize}

\vspace{0.5cm}

SCENARIO 20
Choose your preferred option
\begin{itemize}
    \item OPTION A: Receive £0.50 for sure. And if the virtual six-sided die rolls 5 or 6, decode 20 sequences; otherwise, decode 0 sequences
    \item OPTION B: Decode 0 sequences for £0.25
\end{itemize}

\vspace{0.5cm}

SCENARIO 21
Choose your preferred option:
\begin{itemize}
    \item OPTION A: Receive £0.70 for sure. And if the virtual six-sided die rolls 1, 2, 3 or 4, decode 20 sequences; otherwise, decode 0 sequences
    \item OPTION B: Receive £0.50 for sure. And if the virtual six-sided die rolls 3 or 4 decode 20 sequences; otherwise, decode 0 sequences
\end{itemize}

\subsection{Main task - Example of Simultaneous choices}\label{appendix:screensimultaneous}

\includegraphics[width=0.9\textwidth]{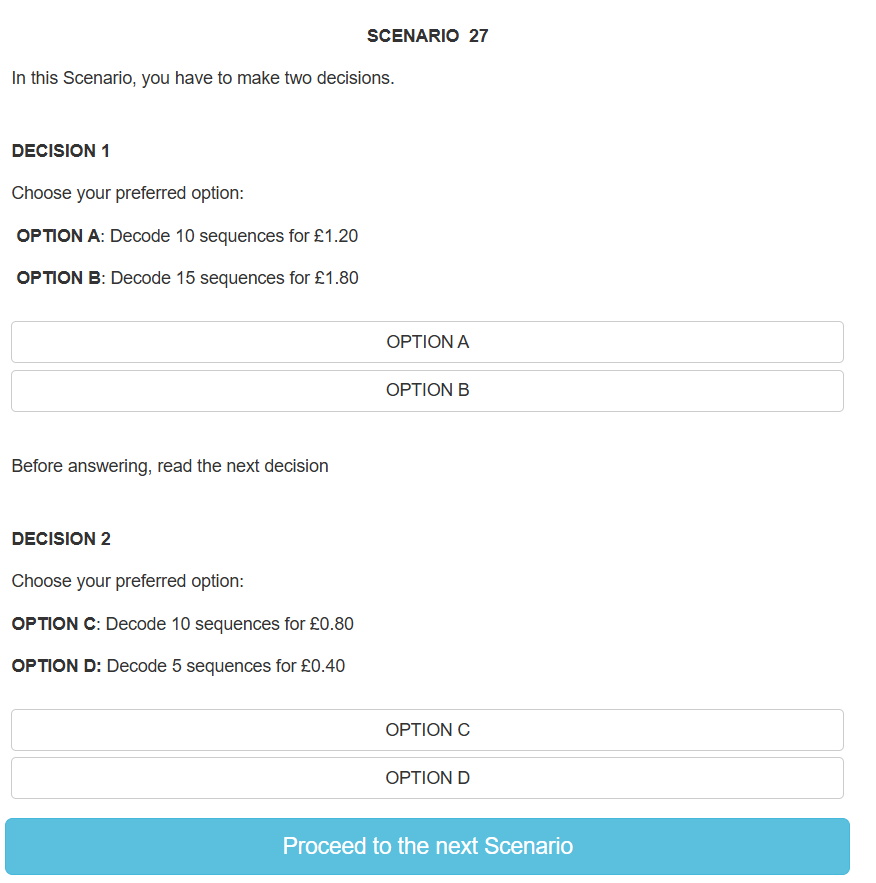}

%% file: appendix-extra-treatments.tex
\pagebreak
\section{Follow-up to Study 1: an Attempt to Debias}\label{appendix:before-after}
In a follow-up study we explored if we can reduce narrow bracketing by making the increasing costs of the additional tasks more salient through a different presentation of choices. We design two new treatments, BEFORE and AFTER, that are identical to the BOTH treatment, except for describing additional sequences as "additional sequences before" or "additional sequences after" the 15 required tasks.

\pagebreak

\textbf{[BEFORE - Scenario 1]}

\textbf{Note: you are required to decode 15 sequences correctly, in addition to the sequences based on your choices.}\\
\textbf{Choices to make now:} for each choice in this Scenario, choose the preferred option.\\
\textbf{By completing the HIT you will receive \$2.00 plus a bonus depending on your choices.}\\

\begin{center}\small{
\flushleft{\begin{tabular}{p{0.2cm}p{8cm}p{0.1cm}p{0.1cm}p{8.2cm}}\hline
&& A & B & \\ \hline
1) & {\raggedright 0 additional sequences before the 15 required for an extra \$4.00} & $\bigcirc$ & $\bigcirc$ & {\raggedright 15 additional sequences before the 15 required for an extra \$4.25} \\
2) & {\raggedright 0 additional sequences before the 15 required for an extra \$4.00} & $\bigcirc$ & $\bigcirc$ & {\raggedright 15 additional sequences before the 15 required for an extra \$4.50} \\
3) & {\raggedright 0 additional sequences before the 15 required for an extra \$4.00} & $\bigcirc$ & $\bigcirc$ & {\raggedright 15 additional sequences before the 15 required for an extra \$4.75} \\
4) & {\raggedright 0 additional sequences before the 15 required for an extra \$4.00} & $\bigcirc$ & $\bigcirc$ & {\raggedright 15 additional sequences before the 15 required for an extra \$5.00} \\
5) & {\raggedright 0 additional sequences before the 15 required for an extra \$4.00} & $\bigcirc$ & $\bigcirc$ & {\raggedright 15 additional sequences before the 15 required for an extra \$5.25} \\
6) & {\raggedright 0 additional sequences before the 15 required for an extra \$4.00} & $\bigcirc$ & $\bigcirc$ & {\raggedright 15 additional sequences before the 15 required for an extra \$5.50} \\
7) & {\raggedright 0 additional sequences before the 15 required for an extra \$4.00} & $\bigcirc$ & $\bigcirc$ & {\raggedright 15 additional sequences before the 15 required for an extra \$5.75} \\
8) & {\raggedright 0 additional sequences before the 15 required for an extra \$4.00} & $\bigcirc$ & $\bigcirc$ & {\raggedright 15 additional sequences before the 15 required for an extra \$6.00} \\
9) & {\raggedright 0 additional sequences before the 15 required for an extra \$4.00} & $\bigcirc$ & $\bigcirc$ & {\raggedright 15 additional sequences before the 15 required for an extra \$6.25} \\
10) & {\raggedright 0 additional sequences before the 15 required for an extra \$4.00} & $\bigcirc$ & $\bigcirc$ & {\raggedright 15 additional sequences before the 15 required for an extra \$6.50} \\
11) & {\raggedright 0 additional sequences before the 15 required for an extra \$4.00} & $\bigcirc$ & $\bigcirc$ & {\raggedright 15 additional sequences before the 15 required for an extra \$6.75} \\
12) & {\raggedright 0 additional sequences before the 15 required for an extra \$4.00} & $\bigcirc$ & $\bigcirc$ & {\raggedright 15 additional sequences before the 15 required for an extra \$7.00} \\
13) & {\raggedright 0 additional sequences before the 15 required for an extra \$4.00} & $\bigcirc$ & $\bigcirc$ & {\raggedright 15 additional sequences before the 15 required for an extra \$7.25} \\
14) & {\raggedright 0 additional sequences before the 15 required for an extra \$4.00} & $\bigcirc$ & $\bigcirc$ & {\raggedright 15 additional sequences before the 15 required for an extra \$7.50} \\
15) & {\raggedright 0 additional sequences before the 15 required for an extra \$4.00} & $\bigcirc$ & $\bigcirc$ & {\raggedright 15 additional sequences before the 15 required for an extra \$7.75} \\
16) & {\raggedright 0 additional sequences before the 15 required for an extra \$4.00} & $\bigcirc$ & $\bigcirc$ & {\raggedright 15 additional sequences before the 15 required for an extra \$8.00} \\
\hline
\end{tabular}}}
\end{center}

\pagebreak

\textbf{[BEFORE - Scenario 2]}

\textbf{Note: you are required to decode 15 sequences correctly, in addition to the sequences based on your choices.}\\
\textbf{Choices to make now:} for each choice in this Scenario, choose the preferred option.\\
\textbf{By completing the HIT you will receive \$2.00 plus a bonus depending on your choices.}\\

\begin{center}\small{
\flushleft{\begin{tabular}{p{0.1cm}p{8.1cm}p{0.1cm}p{0.1cm}p{8.2cm}}\hline
&& A & B & \\ \hline
1) & {\raggedright 15 additional sequences before the 15 required for an extra \$4.00} & $\bigcirc$ & $\bigcirc$ & {\raggedright 30 additional sequences before the 15 required for an extra \$4.25} \\
2) & {\raggedright 15 additional sequences before the 15 required for an extra \$4.00} & $\bigcirc$ & $\bigcirc$ & {\raggedright 30 additional sequences before the 15 required for an extra \$4.50} \\
3) & {\raggedright 15 additional sequences before the 15 required for an extra \$4.00} & $\bigcirc$ & $\bigcirc$ & {\raggedright 30 additional sequences before the 15 required for an extra \$4.75} \\
4) & {\raggedright 15 additional sequences before the 15 required for an extra \$4.00} & $\bigcirc$ & $\bigcirc$ & {\raggedright 30 additional sequences before the 15 required for an extra \$5.00} \\
5) & {\raggedright 15 additional sequences before the 15 required for an extra \$4.00} & $\bigcirc$ & $\bigcirc$ & {\raggedright 30 additional sequences before the 15 required for an extra \$5.25} \\
6) & {\raggedright 15 additional sequences before the 15 required for an extra \$4.00} & $\bigcirc$ & $\bigcirc$ & {\raggedright 30 additional sequences before the 15 required for an extra \$5.50} \\
7) & {\raggedright 15 additional sequences before the 15 required for an extra \$4.00} & $\bigcirc$ & $\bigcirc$ & {\raggedright 30 additional sequences before the 15 required for an extra \$5.75} \\
8) & {\raggedright 15 additional sequences before the 15 required for an extra \$4.00} & $\bigcirc$ & $\bigcirc$ & {\raggedright 30 additional sequences before the 15 required for an extra \$6.00} \\
9) & {\raggedright 15 additional sequences before the 15 required for an extra \$4.00} & $\bigcirc$ & $\bigcirc$ & {\raggedright 30 additional sequences before the 15 required for an extra \$6.25} \\
10) & {\raggedright 15 additional sequences before the 15 required for an extra \$4.00} & $\bigcirc$ & $\bigcirc$ & {\raggedright 30 additional sequences before the 15 required for an extra \$6.50} \\
11) & {\raggedright 15 additional sequences before the 15 required for an extra \$4.00} & $\bigcirc$ & $\bigcirc$ & {\raggedright 30 additional sequences before the 15 required for an extra \$6.75} \\
12) & {\raggedright 15 additional sequences before the 15 required for an extra \$4.00} & $\bigcirc$ & $\bigcirc$ & {\raggedright 30 additional sequences before the 15 required for an extra \$7.00} \\
13) & {\raggedright 15 additional sequences before the 15 required for an extra \$4.00} & $\bigcirc$ & $\bigcirc$ & {\raggedright 30 additional sequences before the 15 required for an extra \$7.25} \\
14) & {\raggedright 15 additional sequences before the 15 required for an extra \$4.00} & $\bigcirc$ & $\bigcirc$ & {\raggedright 30 additional sequences before the 15 required for an extra \$7.50} \\
15) & {\raggedright 15 additional sequences before the 15 required for an extra \$4.00} & $\bigcirc$ & $\bigcirc$ & {\raggedright 30 additional sequences before the 15 required for an extra \$7.75} \\
16) & {\raggedright 15 additional sequences before the 15 required for an extra \$4.00} & $\bigcirc$ & $\bigcirc$ & {\raggedright 30 additional sequences before the 15 required for an extra \$8.00} \\
\hline
\end{tabular}}}
\end{center}
\pagebreak

\textbf{[AFTER  - Scenario 1]}

\textbf{Note: you are required to decode 15 sequences correctly, in addition to the sequences based on your choices.}\\
\textbf{Choices to make now:} for each choice in this Scenario, choose the preferred option.\\
\textbf{By completing the HIT you will receive \$2.00 plus a bonus depending on your choices.}\\

\begin{center}\small{
\flushleft{\begin{tabular}{p{0.2cm}p{7.8cm}p{0.1cm}p{0.1cm}p{8.2cm}}\hline
&& A & B & \\ \hline
1) & {\raggedright 0 additional sequences after the 15 required for an extra \$4.00} & $\bigcirc$ & $\bigcirc$ & {\raggedright 15 additional sequences after the 15 required for an extra \$4.25} \\
2) & {\raggedright 0 additional sequences after the 15 required for an extra \$4.00} & $\bigcirc$ & $\bigcirc$ & {\raggedright 15 additional sequences after the 15 required for an extra \$4.50} \\
3) & {\raggedright 0 additional sequences after the 15 required for an extra \$4.00} & $\bigcirc$ & $\bigcirc$ & {\raggedright 15 additional sequences after the 15 required for an extra \$4.75} \\
4) & {\raggedright 0 additional sequences after the 15 required for an extra \$4.00} & $\bigcirc$ & $\bigcirc$ & {\raggedright 15 additional sequences after the 15 required for an extra \$5.00} \\
5) & {\raggedright 0 additional sequences after the 15 required for an extra \$4.00} & $\bigcirc$ & $\bigcirc$ & {\raggedright 15 additional sequences after the 15 required for an extra \$5.25} \\
6) & {\raggedright 0 additional sequences after the 15 required for an extra \$4.00} & $\bigcirc$ & $\bigcirc$ & {\raggedright 15 additional sequences after the 15 required for an extra \$5.50} \\
7) & {\raggedright 0 additional sequences after the 15 required for an extra \$4.00} & $\bigcirc$ & $\bigcirc$ & {\raggedright 15 additional sequences after the 15 required for an extra \$5.75} \\
8) & {\raggedright 0 additional sequences after the 15 required for an extra \$4.00} & $\bigcirc$ & $\bigcirc$ & {\raggedright 15 additional sequences after the 15 required for an extra \$6.00} \\
9) & {\raggedright 0 additional sequences after the 15 required for an extra \$4.00} & $\bigcirc$ & $\bigcirc$ & {\raggedright 15 additional sequences after the 15 required for an extra \$6.25} \\
10) & {\raggedright 0 additional sequences after the 15 required for an extra \$4.00} & $\bigcirc$ & $\bigcirc$ & {\raggedright 15 additional sequences after the 15 required for an extra \$6.50} \\
11) & {\raggedright 0 additional sequences after the 15 required for an extra \$4.00} & $\bigcirc$ & $\bigcirc$ & {\raggedright 15 additional sequences after the 15 required for an extra \$6.75} \\
12) & {\raggedright 0 additional sequences after the 15 required for an extra \$4.00} & $\bigcirc$ & $\bigcirc$ & {\raggedright 15 additional sequences after the 15 required for an extra \$7.00} \\
13) & {\raggedright 0 additional sequences after the 15 required for an extra \$4.00} & $\bigcirc$ & $\bigcirc$ & {\raggedright 15 additional sequences after the 15 required for an extra \$7.25} \\
14) & {\raggedright 0 additional sequences after the 15 required for an extra \$4.00} & $\bigcirc$ & $\bigcirc$ & {\raggedright 15 additional sequences after the 15 required for an extra \$7.50} \\
15) & {\raggedright 0 additional sequences after the 15 required for an extra \$4.00} & $\bigcirc$ & $\bigcirc$ & {\raggedright 15 additional sequences after the 15 required for an extra \$7.75} \\
16) & {\raggedright 0 additional sequences after the 15 required for an extra \$4.00} & $\bigcirc$ & $\bigcirc$ & {\raggedright 15 additional sequences after the 15 required for an extra \$8.00}\\
\hline
\end{tabular}}}
\end{center}

\pagebreak

\textbf{[AFTER  - Scenario 2]}

\textbf{Note: you are required to decode 15 sequences correctly, in addition to the sequences based on your choices.}\\
\textbf{Choices to make now:} for each choice in this Scenario, choose the preferred option.\\
\textbf{By completing the HIT you will receive \$2.00 plus a bonus depending on your choices.}\\

\begin{center}\small{
\flushleft{\begin{tabular}{p{0.1cm}p{8cm}p{0.1cm}p{0.1cm}p{8.2cm}}\hline
&& A & B & \\ \hline
1) & {\raggedright 15 additional sequences after the 15 required for an extra \$4.00} & $\bigcirc$ & $\bigcirc$ & {\raggedright 30 additional sequences after the 15 required for an extra \$4.25} \\
2) & {\raggedright 15 additional sequences after the 15 required for an extra \$4.00} & $\bigcirc$ & $\bigcirc$ & {\raggedright 30 additional sequences after the 15 required for an extra \$4.50} \\
3) & {\raggedright 15 additional sequences after the 15 required for an extra \$4.00} & $\bigcirc$ & $\bigcirc$ & {\raggedright 30 additional sequences after the 15 required for an extra \$4.75} \\
4) & {\raggedright 15 additional sequences after the 15 required for an extra \$4.00} & $\bigcirc$ & $\bigcirc$ & {\raggedright 30 additional sequences after the 15 required for an extra \$5.00} \\
5) & {\raggedright 15 additional sequences after the 15 required for an extra \$4.00} & $\bigcirc$ & $\bigcirc$ & {\raggedright 30 additional sequences after the 15 required for an extra \$5.25} \\
6) & {\raggedright 15 additional sequences after the 15 required for an extra \$4.00} & $\bigcirc$ & $\bigcirc$ & {\raggedright 30 additional sequences after the 15 required for an extra \$5.50} \\
7) & {\raggedright 15 additional sequences after the 15 required for an extra \$4.00} & $\bigcirc$ & $\bigcirc$ & {\raggedright 30 additional sequences after the 15 required for an extra \$5.75} \\
8) & {\raggedright 15 additional sequences after the 15 required for an extra \$4.00} & $\bigcirc$ & $\bigcirc$ & {\raggedright 30 additional sequences after the 15 required for an extra \$6.00} \\
9) & {\raggedright 15 additional sequences after the 15 required for an extra \$4.00} & $\bigcirc$ & $\bigcirc$ & {\raggedright 30 additional sequences after the 15 required for an extra \$6.25} \\
10) & {\raggedright 15 additional sequences after the 15 required for an extra \$4.00} & $\bigcirc$ & $\bigcirc$ & {\raggedright 30 additional sequences after the 15 required for an extra \$6.50} \\
11) & {\raggedright 15 additional sequences after the 15 required for an extra \$4.00} & $\bigcirc$ & $\bigcirc$ & {\raggedright 30 additional sequences after the 15 required for an extra \$6.75} \\
12) & {\raggedright 15 additional sequences after the 15 required for an extra \$4.00} & $\bigcirc$ & $\bigcirc$ & {\raggedright 30 additional sequences after the 15 required for an extra \$7.00} \\
13) & {\raggedright 15 additional sequences after the 15 required for an extra \$4.00} & $\bigcirc$ & $\bigcirc$ & {\raggedright 30 additional sequences after the 15 required for an extra \$7.25} \\
14) & {\raggedright 15 additional sequences after the 15 required for an extra \$4.00} & $\bigcirc$ & $\bigcirc$ & {\raggedright 30 additional sequences after the 15 required for an extra \$7.50} \\
15) & {\raggedright 15 additional sequences after the 15 required for an extra \$4.00} & $\bigcirc$ & $\bigcirc$ & {\raggedright 30 additional sequences after the 15 required for an extra \$7.75} \\
16) & {\raggedright 15 additional sequences after the 15 required for an extra \$4.00} & $\bigcirc$ & $\bigcirc$ & {\raggedright 30 additional sequences after the 15 required for an extra \$8.00} \\
\hline
\end{tabular}}}
\end{center}

\begin{figure}[H]
  \centering
  \includegraphics[scale=0.8]{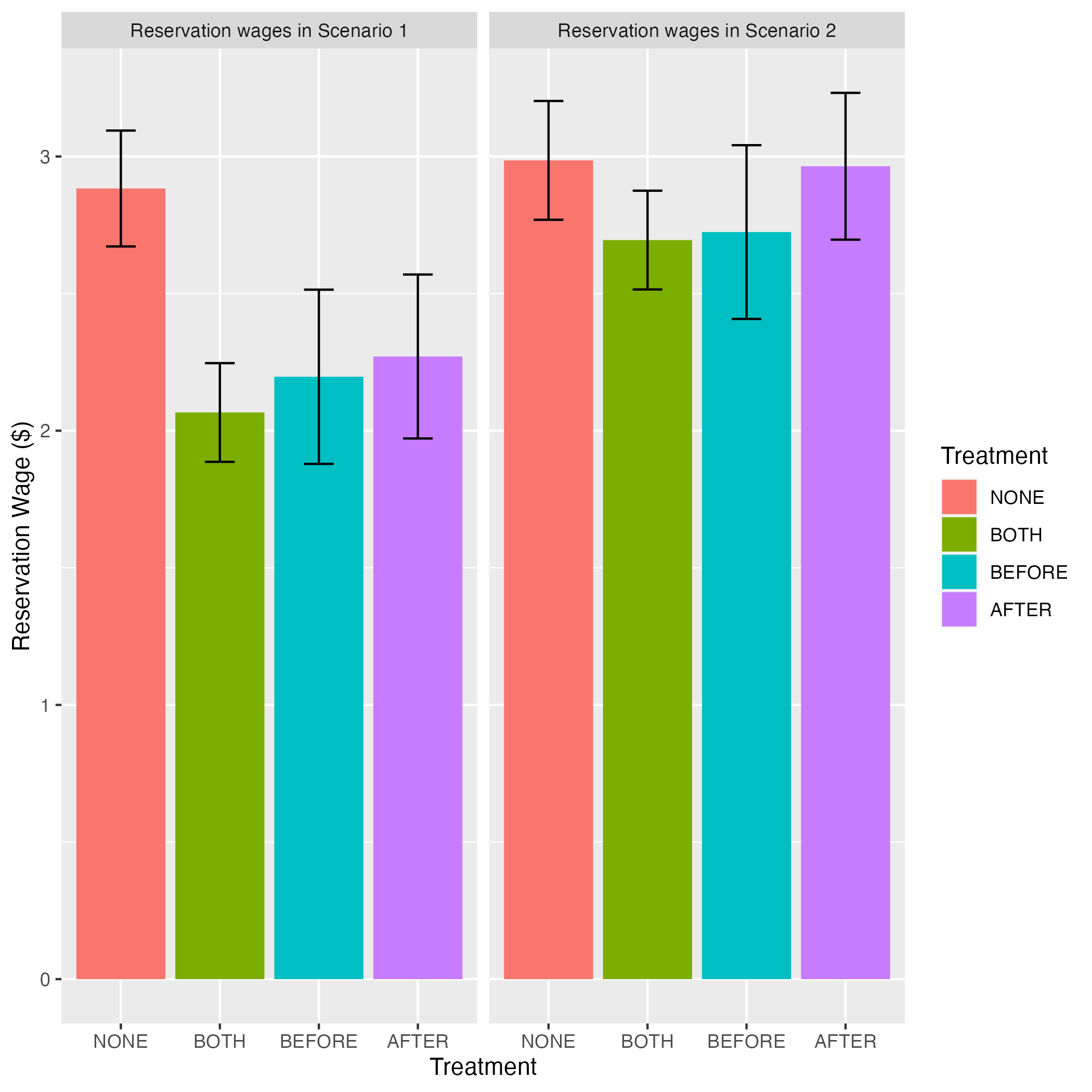}
    \caption{The reservation wages for NONE, BOTH, BEFORE, and AFTER by treatment for each of the two scenarios, along with confidence interval (2 standard errors above and below the estimate).}
    \label{fig:bar-plot-means-ba}
\end{figure}

Figure \ref{fig:bar-plot-means-ba} shows the means by Scenario and by treatment. Treatments BEFORE and AFTER differ from treatment BOTH only by highlighting the number of tasks to do and labeling them as "before" or "after" the baseline tasks. In both BEFORE and AFTER the extra reservation wage is higher than in BOTH, but in both cases this difference is not statistically significant ($p-values$ $>0.097$). In Appendix \ref{appendix:reveal-on-page-only}, we show, however, that the AFTER treatment is statistically significantly different from BOTH when we limit ourselves to those observations in BOTH that received their information about baseline on the first choice page only, which may indicate a partial success of debiasing.

\begin{result}
  We reject the hypothesis that drawing attention to the later tasks reduces narrow bracketing.
\end{result}

\begin{result}
  We find tentative support for hypothesis that drawing attention to convexity by highlighting earlier tasks reduces narrow bracketing.
\end{result}

\subsection{BEFORE and AFTER summary statistics}\label{appendix:before_after_summary}

\import{./data/}{summary_statistics_before_after.tex}

\subsection{Baseline Tasks revealed on the choice page only}\label{appendix:reveal-on-page-only}

Here we report the results when we restrict the data from treatment BOTH to those sessions where the baseline endowment is only revealed on the first choice page, rather than on the page right before, as was inadvertently the case for early sessions.

\import{./data/}{t_test_presentation_on_page.tex}

Broad bracketing is rejected as before, narrow bracketing is still not rejected.

Now BOTH and AFTER are statistically significantly different in this case, as indicated by Table \ref{tab:t_test_before_after_on_page}. However, the issues around the different sample population for BEFORE/AFTER remain, given that we collected most of the data after COVID-19 induced lockdowns.

\import{./data/}{t_test_before_after_on_page.tex}

\pagebreak

%% file: data/summary_statistics_before_after.tex
\begin{table}[H]
\centering
\caption{\label{tab:summary_stats_before_after}Summary statistics for follow-up treatments}
\centering
\fontsize{12}{14}\selectfont
\begin{tabular}[t]{lrrrr}
\toprule
\textbf{} & \textbf{BOTH} & \textbf{BEFORE} & \textbf{AFTER} & \textbf{p-value}\\
\midrule\\
Participants & 320 & 150 & 152 & \\
Attrition & 20.3\% & 18\% & 13.8\% & 0.23\\
Final Participants & 255 & 123 & 131 & \\
\midrule\\
Share Female & 0.38 & 0.36 & 0.39 & 0.68\\
Age & 35 & 35.1 & 35.9 & 0.47\\
Tediousness & 7.45 & 7.48 & 7.56 & 0.93\\
\midrule\\
\addlinespace[0.3em]
\multicolumn{5}{l}{\textbf{Inconsistent Choices}}\\
\hspace{1em}Scenario 1 & 18.8\% & 37.3\% & 33.6\% & 0\\
\hspace{1em}Scenario 2 & 18.4\% & 35.3\% & 34.9\% & 0\\
\bottomrule
\end{tabular}
\end{table}

%% file: data/t_test_presentation_on_page.tex
\begin{table}[H]
\centering
\caption{\label{tab:t_test_on_page}Between-treatment p-values for main treatments based on two-sided t-test, treating each individual in each scenario as a single independent observation. Restricted to those sessions of BOTH where baseline is revealed only on first choice page.}
\centering
\fontsize{12}{14}\selectfont
\begin{tabular}[t]{lcc}
\toprule
Treatments & NONE & BOTH\\
\midrule
\addlinespace[0.3em]
\multicolumn{3}{l}{\textbf{Scenario 1}}\\
\hspace{1em}BOTH & $< 0.001$ & \\
\hspace{1em}MONEY/LOW & $< 0.001$ & $0.106$\\
\addlinespace[0.3em]
\multicolumn{3}{l}{\textbf{Scenario 2}}\\
\hspace{1em}BOTH & $0.006$ & \\
\hspace{1em}MONEY/LOW & $0.120$ & $0.162$\\
\bottomrule
\end{tabular}
\end{table}

%% file: data/t_test_before_after_on_page.tex
\begin{table}
\centering
\caption{\label{tab:t_test_before_after_on_page}Between-treatment p-values for BOTH, BEFORE, AFTER, and NONE treatments based on two-sided t-tests, treating each individual in each scenario as a single independent observation. Restricted to sessions of BOTH when the endowment is mentioned on choice page first.}
\centering
\fontsize{12}{14}\selectfont
\begin{tabular}[t]{lccc}
\toprule
Treatments & BOTH & BEFORE & AFTER\\
\midrule
\addlinespace[0.3em]
\multicolumn{4}{l}{\textbf{Scenario 1}}\\
\hspace{1em}BEFORE & $0.373$ &  & \\
\hspace{1em}AFTER & $0.202$ & $0.735$ & \\
\hspace{1em}NONE & $< 0.001$ & $< 0.001$ & $0.001$\\
\addlinespace[0.3em]
\multicolumn{4}{l}{\textbf{Scenario 2}}\\
\hspace{1em}BEFORE & $0.263$ &  & \\
\hspace{1em}AFTER & $0.015$ & $0.249$ & \\
\hspace{1em}NONE & $0.006$ & $0.175$ & $0.901$\\
\bottomrule
\end{tabular}
\end{table}